\documentclass[10pt,emulateapj,apj]{emulateapj}
\usepackage{color}
\shorttitle{Topology of Galaxy Distribution}
\shortauthors{Choi et al.}
\begin{document}
\newcommand{\hmpc}{{h^{-1}\,{\rm Mpc}}}

\title{Galaxy Clustering Topology in the Sloan Digital Sky Survey
Main Galaxy Sample: a Test for Galaxy Formation Models}

\author{
Yun-Young Choi\altaffilmark{1},
Changbom Park\altaffilmark{2},
Juhan Kim\altaffilmark{1},
J. Richard Gott III\altaffilmark{3},
David H. Weinberg\altaffilmark{4},
Michael S. Vogeley\altaffilmark{5},
and Sungsoo S. Kim\altaffilmark{1,6}\\
(for the SDSS Collaboration)}
\altaffiltext{1}{Dept. of Astronomy \& Space Science, Kyung Hee University,
Gyeonggi 446-701, Korea}
\altaffiltext{2}{Korea Institute for Advanced Study, Dongdaemun-gu, Seoul
130-722, Korea}
\altaffiltext{3}{Department of Astrophysical Sciences, Peyton Hall, Princeton
University, Princeton, NJ 08544-1001, USA}
\altaffiltext{4}{Department of Astronomy and CCAPP, Ohio State University, 
Columbus, OH 43210.}
\altaffiltext{5}{Department of Physics, Drexel University, 3141 Chestnut Street,
Philadelphia, PA 19104, USA}
\altaffiltext{6}{School of Space Research, Kyung Hee University, Yongin,
Gyeonggi 446-701, Korea}

\begin{abstract}
We measure the topology of the main galaxy distribution using the Seventh Data 
Release of the Sloan Digital Sky Survey, examining the dependence of 
galaxy clustering topology on galaxy properties. The observational results are 
used to test galaxy formation models.  A volume-limited sample defined by 
$M_r<-20.19$ enables us to measure the genus curve with amplitude of 
$G=378$ at $6h^{-1}$Mpc smoothing scale, with 4.8\% uncertainty including all 
systematics and cosmic variance.
The clustering topology over the smoothing length interval from
6 to $10 h^{-1}$Mpc reveals a mild scale-dependence for
the shift ($\Delta\nu$) and 
void abundance ($A_V$) parameters of the genus curve.
We find substantial 
bias in the topology of galaxy clustering with respect to the 
predicted topology of the matter distribution, which varies with
luminosity, morphology, color, and the smoothing scale of the density field.
The distribution of relatively brighter galaxies 
shows a greater prevalence of isolated clusters and more percolated voids.  
Even though early (late)-type galaxies show topology similar to
that of red (blue) galaxies, the morphology dependence of topology is not
identical to the color dependence.  In particular, the void abundance 
parameter $A_V$ depends on morphology more strongly than on color.
We test five galaxy assignment schemes applied to cosmological N-body 
simulations of a $\Lambda$CDM universe 
to generate mock galaxies: the Halo-Galaxy one-to-one 
Correspondence model, the Halo Occupation Distribution model, 
and three implementations of Semi-Analytic Models (SAMs).  None of the models
reproduces all aspects of the observed clustering topology; the deviations
vary from one model to another but include statistically significant 
discrepancies in the abundance of isolated voids or isolated clusters and
the amplitude and overall shift of the genus curve.  SAM predictions of the
topology color-dependence are usually correct in sign but incorrect in
magnitude.
Our topology tests indicate that, in these models, voids should be emptier 
and more connected, and the threshold for galaxy formation should be at lower
densities.

\end{abstract}


\section{Introduction}
Galaxy clustering has long been used to constrain cosmological models and to 
understand formation of galaxies. The most extensively studied clustering 
statistics are the autocorrelation function and the power spectrum. 
They are Fourier transforms of each other, and measure the clustering strength
as a function of scale. The ``Standard Cold Dark Matter'' model (with
the density parameter $\Omega_m=1$, scale-invariant primordial fluctuations, 
and standard relativistic particle background), popular in the 1980s, 
was ruled out by showing that it was inconsistent with the observed 
galaxy correlation function (Maddox et al. 1990) and power spectrum 
(Vogeley et al. 1992; Park et al. 1992, 1994). 
These statistics are also used to measure the biasing in the galaxy clustering 
amplitude with respect to matter, and are key constraints on galaxy formation 
models connecting dark matter halos and luminous galaxies such as 
semi-analytic models (SAM; e.g., White \& Frenk 1991; Kauffmann et al.\ 1993;
Cole et al. 1994, 2000; Benson et al. 2002; Bower et al. 2006;
Cattaneo et al. 2006; Croton et al. 2006; De Lucia \& Blaizot 2007; 
Monaco et al. 2007; Somerville et al. 2008) 
and halo occupation distribution models (e.g., Seljak 2000; 
Peacock \& Smith 2000; Scoccimarro et al. 2001; Berlind \& Weinberg 2002; 
Kang et al. 2002; Berlind et al. 2003; Zheng, Coil, \& Zehavi 2007).

Topology analysis was introduced by Gott et al. (1986, 1987) to test the 
Gaussianity of the primordial density fluctuations, which is one of the key 
characteristics of simple inflationary models (Bardeen et al. 1986). 
At large scales density fluctuations are still in the linear regime 
and maintain their initial topology, and it is possible to check whether 
or not the primordial fluctuations were a Gaussian field. 
At smaller scales, non-linear gravitational evolution and biased 
galaxy formation make the topology of the observed galaxy distribution 
deviate from the Gaussian form even if the initial conditions were Gaussian 
distributed as shown by perturbation theories and large N-body 
simulations (Park \& Gott 1991; Weinberg \& Cole 1992; Matsubara 1994; 
Matsubara \& Suto 1996); 
using fractional volume rather than density threshold as
the independent variable in topology analysis mitigates but does not
eliminate these non-linear and biasing effects (Weinberg et al.\ 1987;
Melott et al.\ 1988).
Through studies of many observational samples, the topological
properties of the large-scale distribution of galaxies have been 
examined (Gott et al. 1989; 
Park, Gott \& da Costa 1992; Moore et al. 1992; 
Vogeley et al. 1994; Rhoads, Gott, \& Postman 1994;
Protogeros \& Weinberg 1997; Canavezes et al. 1998;
Park, Gott \& Choi 2001; Hoyle, Vogeley \& Gott 2002;
Hikage et al. 2002, 2003; 
Park et al. 2005; James, Lewis \& Colless 2007; Gott, Choi \& Park 2009;
James et al. 2009).
On non-linear or quasi-linear scales, topology analysis is 
useful in constraining both cosmological parameters and galaxy formation 
mechanisms (Park, Kim \& Gott 2005; Gott et al. 2008). 
In particular, differences in clustering topology for different types of 
galaxies reflect their different history of formation and evolution. 
Therefore, looking at the topology of large-scale structure traced by 
different types of galaxies can put strong constraints on galaxy formation 
mechanisms. 

In this paper we use the Seventh Data Release (DR7; Abazajian et al. 2009)
of the Sloan Digital Sky Survey (York et al.\ 2000) to measure the topology
of the galaxy distribution and its dependence on galaxy luminosity, color,
and morphology.  DR7 constitutes the final release of the SDSS Legacy Survey,
and thus of the main SDSS galaxy redshift survey.  We supplement the SDSS data 
with missing redshifts to increase the completeness of the redshift catalog. 
We then generate a set of volume-limited samples of the SDSS galaxies 
divided according to their luminosity, morphology, and color to study the 
relation between the topology and properties of galaxies tracing the 
large-scale structure.

\section{The KIAS-VAGC catalog}

We prepare a catalog containing 593,514 redshifts of the SDSS (York et al. 2002;
Stoughton et al. 2002) main galaxies (Strauss et al. 2002) in 
the magnitude range of $10 < r \leq 17.6$, which is called the Korea Institute
for Advanced Study (KIAS) Value-Added Galaxy Catalog (VAGC) (Choi et al. 2010).
Its main source is the New York University VAGC Large Scale Structure 
Sample ({\tt brvoid0}) from Blanton et al. (2005), which lists redshifts of 
583,946 galaxies with $10 < r \leq 17.6$ (8,562 with $10 < r \leq 14.5$). 
We excluded 929 objects that are in
error. They are outer parts of large galaxies that are deblended by the 
automated photometric pipeline, blank fields, or stars, etc.
Out of the remaining 583,017 galaxies, 5.7\% of galaxies 
in the NYU VAGC LSS do not have spectra due to fiber collision and have 
redshifts borrowed from the nearest neighbor galaxy.
We added redshifts of 10,497 galaxies with $10 < r \leq 17.6$ (1,455 
with $10 < r \leq 14.5$) to the KIAS VAGC that are missing in the NYU VAGC LSS. 
An SDSS photometric catalog and various existing redshift catalogs such as 
the Updated Zwicky Catalog (UZC; Falco et al. 1999), 
the {\it IRAS} Point Source Catalog Redshift Survey (PSCz; Saunders et al. 2000), 
the Third Reference Catalogue of Bright Galaxies (RC3; de Vaucouleurs et al. 1991), 
and the Two Degree Field Galaxy Redshift Survey (2dFGRS; Colless et al. 2001) 
are used in this step.  During this supplementation process,
we corrected positions of galaxies having wrong central positions, separated 
merging objects, and removed spurious objects such as parts of big late-type 
galaxies.  The angular survey mask 
given by NYU-VAGC is maintained and covers 2.562 sr of sky with 
angular selection function greater than 0. 
While the SDSS main galaxy sample extends to $r=17.77$ over most of the
sky, we retain the shallower limit $r=17.60$ used during the early part of
the survey (see Strauss et al.\ 2002) to maintain a homogeneous depth
of the survey volume and thus a less complicated outer boundary.
Extending to $r=17.77$ would add roughly $114,303$ galaxies (from 
NYU VAGC LSS {\tt full0} sample).

\begin{figure*}
\epsscale{0.9}
\plotone{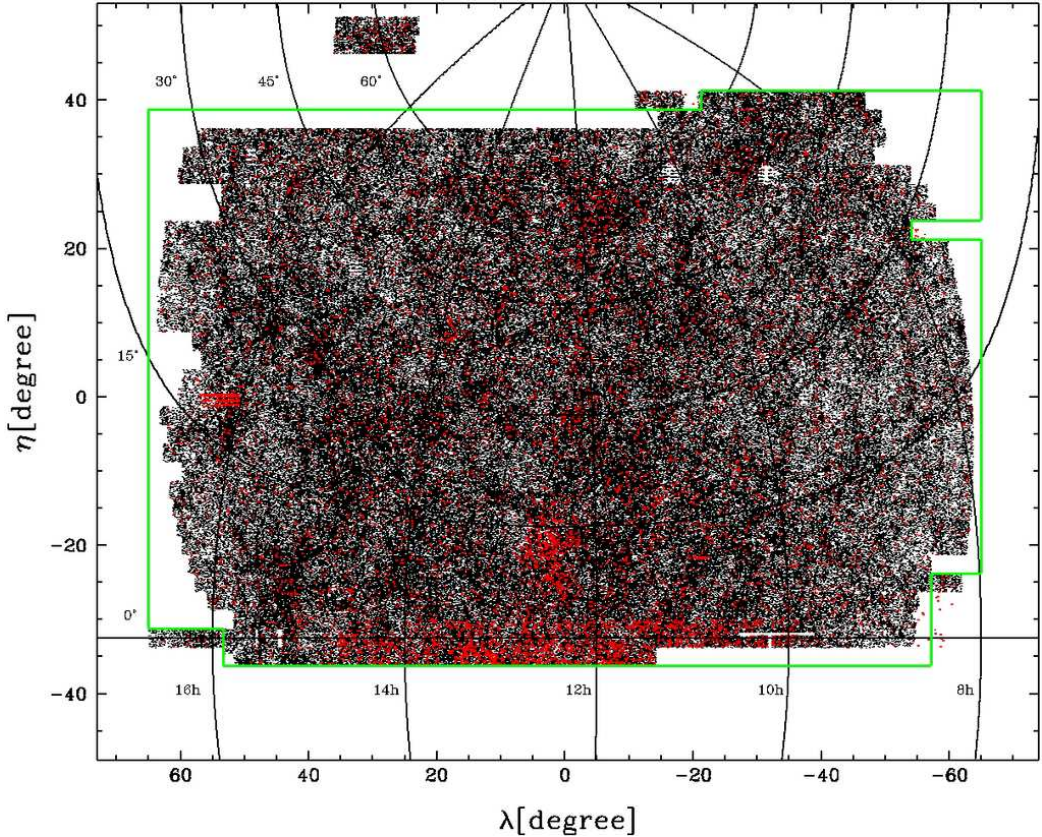}
\caption{ 
SDSS galaxies in the northern galactic hemisphere. 
Black points are the galaxies 
in the NYU VAGC LSS catalog 
and red points are those newly added in the KIAS-VAGC. Green solid 
lines delineate the boundaries of the analysis regions. 
$(\eta,\lambda)$ are the SDSS survey coordinates. Lines of constant RA and DEC
lines are shown in black as roughly vertical and horizontal curves, respectively.} 
\label{survey}
\end{figure*}

The resulting KIAS VAGC has an angular selection function significantly better 
than the original NYU VAGC LSS sample in high surface density regions.
Figure~\ref{survey} shows the distribution of the SDSS galaxies in the northern galactic 
hemisphere. To minimize boundary effects and to keep the surface-to-volume
ratio low in our topology analysis, 
we do not use the galaxies in three narrow stripes of the southern Galactic cap 
and in the small region containing the Hubble Deep Field, 
and we trimmed some regions with narrow angular extent.
What remains are the galaxies inside the green solid boundary shown in 
Figure~\ref{survey}.
The remaining survey region with the angular selection function greater than 0 
covers 2.33 str. 
Black points in Figure~\ref{survey} are the galaxies in NYU VAGC LSS catalog 
and red points are those newly added. 
It can be seen that the added galaxies are mostly 
located in high surface density regions. In particular, a large number of 
galaxies are added in the Virgo cluster region and near the equator. 
Correspondingly, we recalculate the angular selection function of the KIAS 
VAGC within each spherical polygon formed by the adaptive tiling algorithm 
(Blanton et al. 2003) used for the SDSS spectroscopy. Choi et al. (2010) show
the angular selection function before and after supplementation.
After the supplementation, the survey area (in the northern hemisphere)
having angular selection function greater than 0.97 increases
from 39.8\% to 54.3\% of the area with the selection function greater 
than 0. 

All atlas images of galaxies in our KIAS VAGC are downloaded 
from Princeton SDSS reduction server\footnote{http://photo.astro.princeton.edu}
and analyzed to measure 
the seeing- and inclination-corrected $i$-band Petrosian radius, 
$i$-band inverse concentration index, $c_{\rm in}$ and 
$\Delta(g-i)$ color gradient (Park \& Choi 2005; 
Choi et al. 2007). 
These parameters are included in KIAS VAGC together with other fundamental 
photometric and spectroscopic parameters supplied by the NYU VAGC. 

All galaxies in KIAS VAGC are classified into early (elliptical and lenticular) 
and late (spiral and irregular) morphological types. 
We use the automated classification scheme developed by 
Park \& Choi (2005) using $u-r$ color, $\Delta(g-i)$ color gradient, 
and the inverse concentration index $c_{\rm in}$. Reliability and 
completeness of this classification reach about 90\%. To improve the 
automated classification results by visual inspection we selected 82,323 
galaxies that are located in the ``trouble zones'' where the reliability 
of the automated classification is low. These are the galaxies having 
neighbors at very small separations (for these $c_{\rm in}$ is inaccurate) 
or those classified as blue early types or red late types. Thirteen 
astronomers made the visual inspection to check the morphological type of 
these galaxies. As a result, the morphology of 7\% of the inspected galaxies 
was changed and some spurious objects are removed. We kept the morphology of 
5,956 galaxies in the SDSS LSS-DR4plus sample that were already inspected 
by Choi et al. (2007). 

\begin{figure*}
\epsscale{0.8}
\plotone{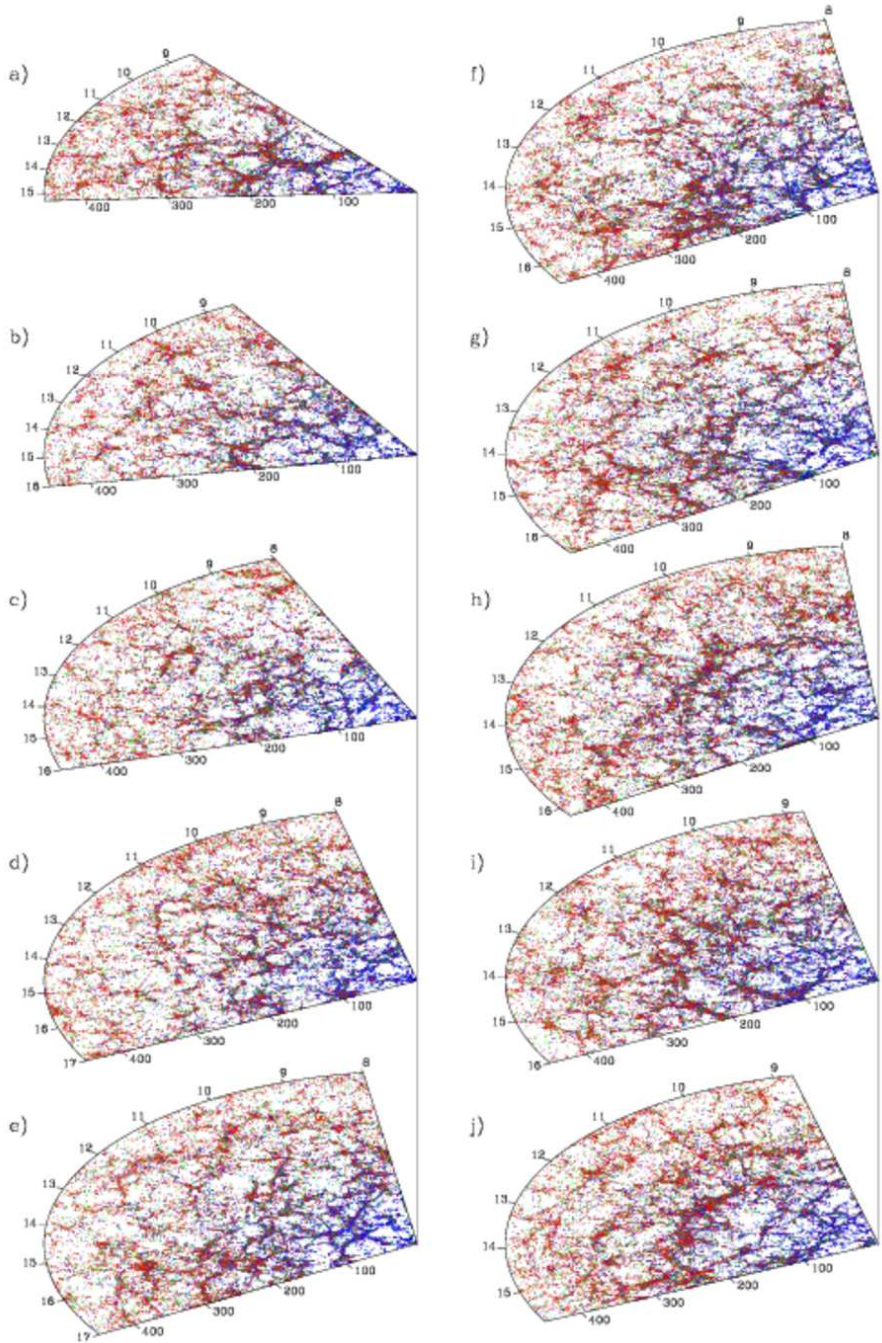}
\caption{Distribution of SDSS galaxies in the legacy survey area
divided into ten wedge slices. Each diagram has radius of 450 $h^{-1}$Mpc 
and thickness of 6 degrees in declination. The boundaries of each wedge diagram
are defined in Table~\ref{tab1}.
Galaxies are color-coded by $u-r$ color. Blue, green, and red dots represent 
galaxies with $u-r<2.0$, $2.0\leq u-r <2.5$, and $u-r\ge 2.5$, respectively.
The radial coordinate is the comoving distance in units of $h^{-1}$Mpc.  
Right ascension is given in units of hour.
A part of the Sloan Great Wall (Gott et al. 2005) is seen in slice {\it j}.}
\label{wedge}
\end{figure*}

Figure 2 shows the galaxies of the KIAS-VAGC contained in ten consecutive
wedge slices in the main survey area.
Each slice has the radius of 450 $h^{-1}$Mpc and is 6 degree thick in
declination. The range in right ascension is varied according to the survey
boundaries shown in Figure 1. The right ascension and declination ranges and
the number of galaxies in each slice are listed in Table 1.
Galaxies are distinguished according to
their color: blue is given to those with $u-r<2.0$, green is for
$2.0\le u-r <2.5$, and red is for $u-r\ge 2.5$. Since we are showing
galaxies in the apparent magnitude-limited sample, galaxies are on average
faint and blue at small distances and bright and red at large distances.
A part of the Sloan Great Wall (Gott et al. 2005) is seen in the bottom slice (slice {\it j},
at $r \approx 250 h^{-1}\,$Mpc),
and the CfA Great Wall (Geller \& Huchra 1989) appears in slice {\it e}
at $r \approx 100 h^{-1}\,$Mpc.
A part of the Cosmic Runner (Park et al. 2005) shows up in the top slice {\it a}. 
There are also numerous prominent superclusters, notably a group of superclusters
in slice {\it f} and a few rod-shaped ones in slice {\it i}.
In addition to these large scale overdense structures, many large 
voids are prevalent in the survey volume.
It should be also noted that none of these features really stands out any more;
the Sloan Great Wall is just a high end of the distribution of structures that are
quite common. A weak circular symmetry around the observer is seen in slice {\it j},
but we do not have such impression in other slices.

\begin{figure}
\plotone{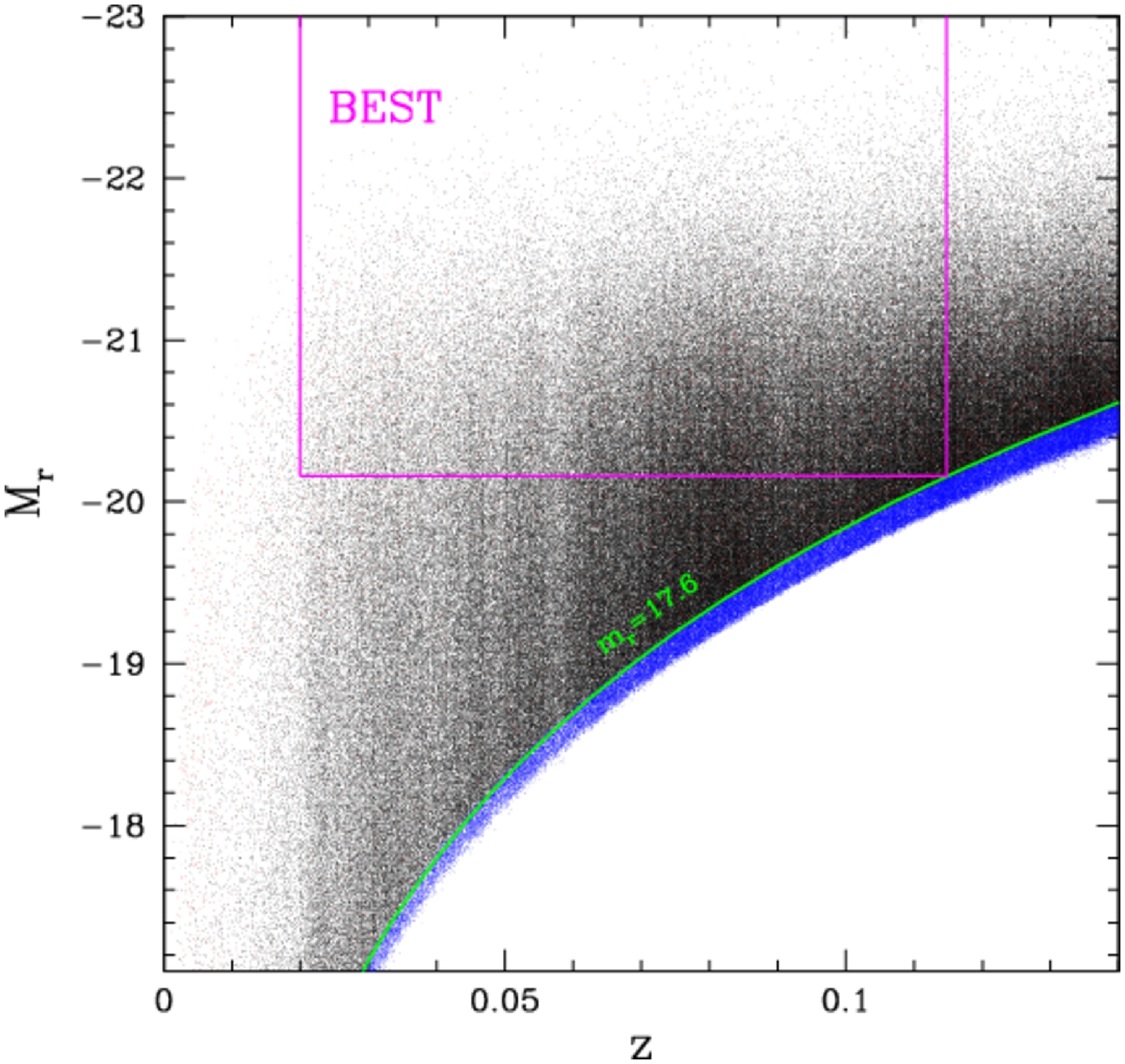}
\caption{Galaxies in the KIAS VAGC shown in $M_r$-$z$ space.
The green curve is the flux limit of $r=17.6$,
and the straight lines in magenta define the sample BEST, which includes
the maximum number of galaxies among volume-limited samples.
Black points are the galaxies in the NYU VAGC LSS catalog and
red points are those newly added in the KIAS-VAGC. 
Blue points are those to be added when the limiting apparent magnitude
extends to $r= 17.77$, but they are not used in this study. 
}
\label{Mz}
\end{figure}

We will use several volume-limited samples defined by $r$-band absolute 
magnitude cuts and the redshift limits of $z_{\rm min}=0.02$ and $z_{\rm max}$, 
where $z_{\rm max}$ is determined by the survey flux limit $r=17.6$ and the 
absolute magnitude cut. 
We frequently refer to the ``mean separation'' between galaxies in a sample,
by which we mean $\bar{d} \equiv n_g^{-1/3}$ where $n_g$ is the number density
of galaxies.
Figure~\ref{Mz} shows the galaxies in our catalog in the 
redshift versus $r$-band absolute magnitude ($M_r$) plane 
and the boundary lines defining one of our 
volume-limited samples ``BEST'', which contains the maximum number 
of galaxies (133,947) among volume-limited samples. 
It is amusing to note that the BEST sample is about 1000 times
larger than the galaxy sample (drawn from the CfA1 redshift survey)
first used for topological analysis by
Gott et al.\ (1986, 1987).

The WMAP 3 year cosmological parameters of
$\Omega_m=0.27, \Omega_{\Lambda}=0.73, \Omega_b = 0.044,
h=0.72, n{_s}=0.958$ (Spergel et al. 2007) are used to relate redshift and 
comoving distance.

\begin{table}
\begin{center}
\caption{Survey Regions for the Wedge Diagrams in Figure~\ref{wedge}}
\label{tab:sim}
\begin{tabular}{ccccc} 
\hline\hline
\noalign{\smallskip}
Panel Name&Right Ascension& Declination &Num. of galaxies \\
\hline
{\it a} &$ 8.5^\circ - 15.3^\circ$ &$54^\circ - 60^\circ$ &22,237\\
{\it b} &$ 8.5^\circ - 16.0^\circ$ &$48^\circ - 54^\circ$ &23,950\\
{\it c} &$ 8.0^\circ - 16.0^\circ$ &$42^\circ - 48^\circ$ &27,232\\
{\it d} &$ 8.0^\circ - 17.0^\circ$ &$36^\circ - 42^\circ$ &36,962\\
{\it e} &$ 8.0^\circ - 17.0^\circ$ &$30^\circ - 36^\circ$ &41,217\\
{\it f} &$ 8.0^\circ - 16.5^\circ$ &$24^\circ - 30^\circ$ &47,140\\
{\it g} &$ 8.0^\circ - 16.5^\circ$ &$18^\circ - 24^\circ$ &45,728\\
{\it h} &$ 8.0^\circ - 16.2^\circ$ &$12^\circ - 18^\circ$ &47,036\\
{\it i} &$ 8.8^\circ - 16.0^\circ$ &$~~   6^\circ - 12^\circ$ &45,554\\
{\it j} &$ 8.8^\circ - 15.7^\circ$ &$~~   0^\circ - ~ 6^\circ$ &43,088\\
\noalign{\smallskip}
\hline
\end{tabular}
\end{center}
\end{table}

\section{The genus statistic}
We measure the topology of the galaxy distribution using the general
methods set out by Gott et al.\ (1987, 1989), with statistical 
characterizations of genus curve shape
(Eqs.~\ref{eq:dnu} and~\ref{eq:avac} below) introduced by Park et al. (1992, 2005).
The point distribution of galaxies is smoothed by a constant-size
Gaussian filter and 
iso-density contour surfaces of the smoothed galaxy density distribution
are searched to calculate the genus. 
The genus is defined as
\begin{equation}
G = {\rm Number~of~holes} -{\rm Number~of~isolated~regions}
\end{equation} 
in the iso-density contour surfaces at a given threshold level, $\nu$,
which is related to the volume fraction $f$ on
the high density side of the density contour surface by
\begin{equation}
f = {1\over\sqrt{2\pi}}\int_\nu^\infty e^{-x^2/2} \,dx.
\end{equation}
The $f = 50\%$ contour
corresponds to the median volume fraction contour ($\nu = 0$).
In our analysis we will always use the volume fraction to find the contour 
surfaces and the threshold level $\nu$ is related with the volume fraction 
by Equation 2.

For Gaussian random phase initial conditions the genus curve is:
\begin{equation}
\label{eq:GRgenus}
g(\nu) = A(1 - \nu^2)e^{-\nu^2/2},
\end{equation}
where the amplitude $A = (\langle k^2 \rangle /3)^{3/2}/2\pi^2$
and $\langle k^2 \rangle$ is the average value of $k^2$ in the
smooth power spectrum
(Hamilton, Gott, \& Weinberg 1986; Doroshkevich 1970).
Deviation of an observed genus curve from Equation~\ref{eq:GRgenus}
indicates the non-Gaussian signal.


The shape of the genus curve can be parameterized by several variables
(Park et al. 1992, 2005; Vogeley et al. 1994).
The first is the amplitude of the genus curve as measured by the
amplitude of the best fitting Gaussian curve of 
Equation~\ref{eq:GRgenus}. 
This gives information about the power spectrum and phase correlation
of the density fluctuation.  Deviations in the shape of the genus
curve from the theoretical random phase case can be quantified
by the following three variables:
\begin{equation}
\label{eq:dnu}
\Delta\nu = \frac{\int_{-1}^1 g(\nu)\nu\,d\nu}{\int_{-1}^1
  g_{\rm G}(\nu)\,d\nu}, ~~~
\end{equation}
\begin{equation}
\label{eq:avac}
A_V = \frac{\int_{-2.2}^{-1.2}g(\nu)\,d\nu}
{\int_{-2.2}^{-1.2}g_{\rm G}(\nu)\,d\nu}, ~~~
A_C = \frac{\int_{1.2}^{2.2}g(\nu)\,d\nu}
{\int_{1.2}^{2.2}g_{\rm G}(\nu)\,d\nu},
\end{equation}
where $g_{\rm G}({\nu})$ is the genus of the best-fit Gaussian curve
(Eq. 3). Thus $\Delta\nu$ measures any shift in the central portion of the
genus curve.  The Gaussian curve (Eq. 2) has $\Delta\nu = 0$.
A negative value of $\Delta\nu$ is frequently called a ``meatball shift'' 
caused by a greater prominence of isolated connected high-density
structures that push the genus curve to the left (Weinberg et al.\ 1987).
$A_V$ and $A_C$ measure the observed number of voids and 
(super) clusters relative to those expected from the best-fitting 
Gaussian curve. 

Because the genus-related statistics are defined as in Equations (4) and
(5), they depend on the amplitude of the best-fit Gaussian curve, which
has a statistical uncertainty. The fluctuation in the amplitude
yields numerical uncertainties in these statistics. One might want to fix
the amplitude using the observed power spectrum and the relation between
the power spectrum and the amplitude of a Gaussian genus curve
(given just after Eq. 3).
But the power spectrum measured from an observational sample also has statistical 
uncertainties, which propagate to the uncertainty in the amplitude of the genus 
curve. Furthermore, the genus versus power spectrum relation in the case of 
Gaussian fields cannot be directly applied to non-Gaussian fields in general.
We also want to separate the non-Gaussianity seen in the genus curve
due to the amplitude drop from that due to the deviation of the genus curve
from the Gaussian shape, and this is best done when the observed genus curve
is compared with a Gaussian curve with their amplitudes matched together.
When galaxy formation models are tested against observations in section 5,
we compare the genus-related statistics calculated in the exactly same way
for both observational data and the mock samples of galaxy formation models.

\section{Results}

\subsection{Genus Measurement} 

\begin{figure}
\plotone{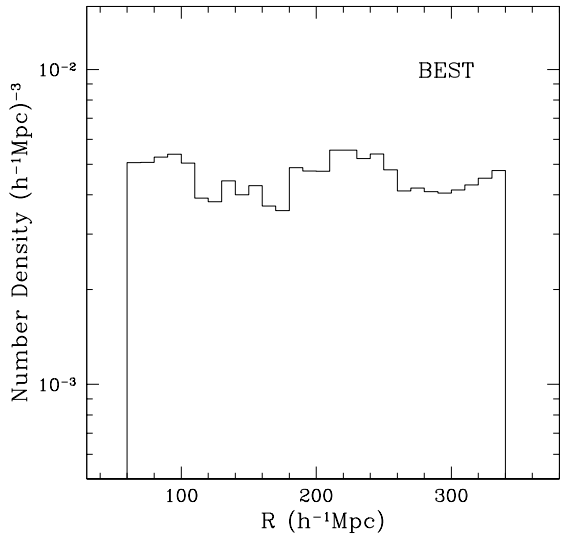}
\caption{Comoving number density of galaxies in the
BEST sample as a function of radial comoving distance $R$.
}
\label{denhis}
\end{figure}

We first study how the topology of large scale structure changes as the 
smoothing scale changes. For this purpose we use the volume-limited sample, 
BEST. 
It includes galaxies with $M_r < -20.19 +5 {\rm log} h$ ($5{\rm log} h$ will be 
subsequently dropped in the expression of $M_r$) and $0.02<z<0.116$, where the 
upper redshift limit is determined by the apparent magnitude limit $r=17.6$ 
and $M_r=-20.19$. 
The mean separation between galaxies in BEST is $\bar{d} = n_g^{-1/3} = 
6.1 h^{-1}$Mpc. 
Figure~\ref{denhis} shows the comoving number 
density of galaxies in BEST as a function of radial comoving distance $R$. 
It can be seen that the comoving density is roughly constant, 
but there is about 20\% excess in the density at $R=210\sim 250 h^{-1}$Mpc 
caused by the Sloan Great Wall.
A genus curve obtained from BEST is shown in Figure~\ref{gplot} 
and the data are given in Table~\ref{tab:genusdcut} in Appendix~\ref{app:genus}.
\begin{figure*}
\epsscale{1.1} 
\plotone{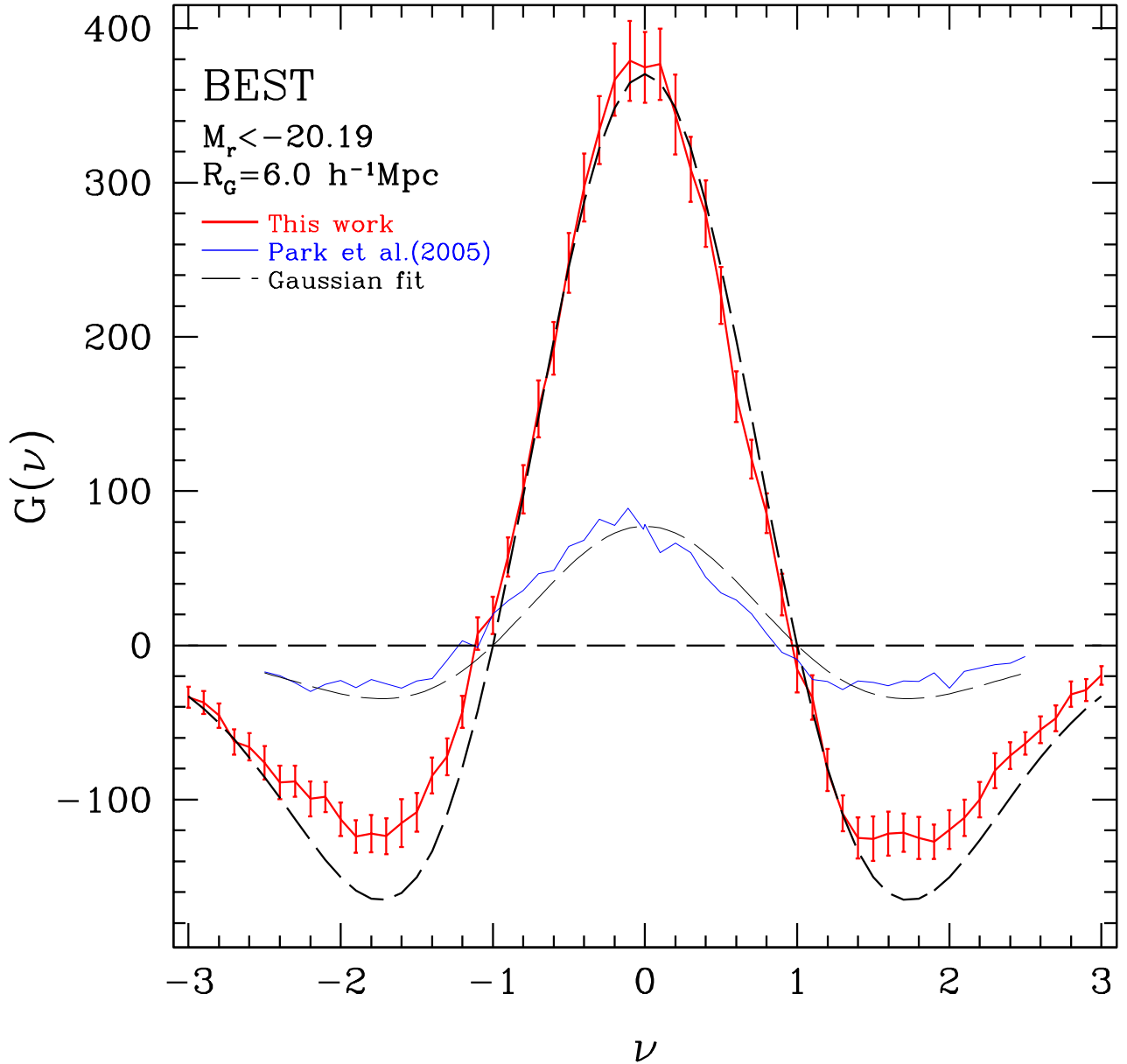}
\caption{A genus curve measured from the BEST sample (solid line with 
error bars), and its best-fit Gaussian curve (long-dashed line). 
Curves with smaller amplitudes were measured from the
BEST sample of SDSS DR3 by Park et al. (2005).
The observed genus curves in this plot have not been corrected for systematic biases. 
} 
\label{gplot}
\end{figure*}

To calculate the genus a smooth density field is obtained 
from galaxy positions in comoving space 
smoothed by an $R_{G} = 6.0 h^{-1}$Mpc Gaussian,
$(2\pi R_{G}^2)^{-3/2} \exp (-r^2/2R_{G}^2)$.
This definition differs from that of some previous works 
(Hamilton et al. 1986; Park et al. 1992; Vogeley et al. 1994), which is related with
current definition by $\lambda=\sqrt{2}R_{G}$.
The solid line with error bars shows measurements from the observational data, 
and the long-dashed line is its best-fit Gaussian curve. As a comparison, we 
include the results (curves with smaller amplitudes) from the BEST sample of 
SDSS Data Release 3 (DR3) measured by Park et al. (2005). As the DR7 sample covers the 
northern sky without a gap, the volume-to-surface ratio increased significantly 
since DR3, and the amplitude of the genus curve increased remarkably. 
The amplitude is now $G=373\pm 18$ with only 4.6\% uncertainty, including all 
systematic effects and cosmic variance.

Figure~\ref{gplot} shows that the observed genus curve deviates from the
Gaussian one in such a way that the genus at high thresholds has smaller
amplitude.
This means there are fewer voids and fewer superclusters when compared with
the Gaussian genus curve.
In other words, voids and superclusters are
more connected and their sizes are larger than those expected for
Gaussian fields.
In addition, the positive ($G>0$) region of the genus curve is
slightly but systematically shifted towards lower $\nu$ values,
and the percolation of large-scale structure (maximum $G$)
occurs slightly below $\nu=0$.
According to the perturbation theory
prediction of Matsubara (1994),
if there were only non-linear gravitational effects involved, with no biasing,
then one should find $A_V + A_C =2$ at all scales (Park et al. 2005),
and the N-body results presented for dark matter in Figure~\ref{sys} below
are consistent with this expectation.
Since the observed $A_V$ and $A_C$ are both less than $1$, it  shows
that biasing effects are involved.

\subsection{Mock Surveys and Systematic Biases}
The genus curve in Figure~\ref{gplot} is affected by peculiar velocity
distortions, boundary effects, and 
finite sampling of the density field.
However, using detailed comparisons to mock catalogs from 
cosmological simulations, we will show that the deviations
from the Gaussian field prediction, and indeed from the expected
non-linear matter density field, are genuine, and statistically 
significant.


We use mock galaxies to estimate uncertainties in the measured genus and
to measure the systematic effects due to survey boundaries, variation of
angular selection function, galaxy biasing, and redshift space distortion.
We adopt the halo-galaxy one-to-one correspondence (HGC) model of
Kim et al. (2008) to assign mock galaxies to dark halos identified in a
cosmological N-body simulation. For this purpose we made two N-body
simulations of the $\Lambda$CDM model having the WMAP 3 year cosmological
parameters.
The simulation and cosmological parameters adopted in the simulations
(S1 and S2) are listed in Table~\ref{tab:sim}. 
\begin{table}
\begin{center}
\caption{N-body Simulation Parameters}
\label{tab:sim}
\begin{tabular}{rrr} 
\hline\hline
\noalign{\smallskip}
Parameter& S1& S2\\
\hline
$N_p\footnote{Number of CDM particles}$&$2048^3$     &$2048^3$\\
$N_m     $&$2048^3$     &$2048^3$\\
$L_{box}  $&$1024\hmpc$  &$1433.6\hmpc$\\
$N_{step}$\footnote{Number of global evolution timesteps from $z_i$ to present epoch}&1880         &1250    \\
$z_i\footnote{Initial redshift where the simulation started}$&47           &50  \\
$m_p\footnote{Simulation particle mass} $    &$8.3\times10^9h^{-1}M_\odot$
 &$2.3\times10^{10}h^{-1}M_\odot$\\
$\epsilon$\footnote{Force resolution}&$0.05\hmpc$           &$0.07\hmpc$ \\
$M_{h,min}\footnote{Minimum halo mass}$    & $2.5\times 10^{11}h^{-1}M_\odot$ & $6.8\times 10^{11}h^{-1}M_\odot$\\
$\bar{d_h}\footnote{Mean halo separation}$  & $4.4 \hmpc$ & $5.9\hmpc$\\
$h       $&0.737        &0.737  \\
$\Omega_m$&0.238        &0.238  \\
$\Omega_b$&0.042        &0.042  \\
$\Omega_{\Lambda}$&0.762 &0.762 \\
$n_s$   &0.958                 &0.958\\
$b   $    &1.314                 &1.314\\
\noalign{\smallskip}
\hline
\end{tabular}
\end{center}
\end{table}
                        
The simulations were started at the initial redshift of $z_{i}$, and
evolved to the present epoch after making $N_{step}$ global timesteps.
Then dark matter particles
at the present epoch are used to identify the dark halos.
The CDM particles in high density regions are first found by using the
Friend-of-Friend (FoF) algorithm with connection length of 0.2 times the mean
particle separation, and then gravitationally bound and tidally stable dark
halos are identified within each FoF particle group (Kim \& Park 2006).
These dark halos include isolated, central and
satellite halos composing FoF halos.
The halos are required to contain at least 30 particles, which corresponds to
$2.5 \times 10^{11} h^{-1}M_{\odot}$ for $1024\hmpc$ box size simulation
(S1) and $6.8 \times 10^{11} h^{-1}M_{\odot}$ for $1433.6\hmpc$ box size one
(S2).
The resulting halo samples include 12,326,725 and 14,244,305 dark matter halos,
and the corresponding mean halo separations are $4.4\hmpc$ and $5.9\hmpc$,
respectively.

To select mock galaxies corresponding to
those in BEST from the S1 simulation,
we sort the dark halos in mass and choose those above a
mass cut so that the resulting halo set has mean halo separation
of $6.1 h^{-1}$Mpc which is the mean separation between galaxies in BEST.
These halos are identified as the galaxies that can be compared with those 
in BEST.

We make 27 mock BEST samples originating at
random locations within the S1 simulation 
using exactly the same survey mask, angular selection function, radial
boundaries, and smoothing length, and also taking into account the peculiar
velocities of the mock galaxies by adding the line of sight component of the 
peculiar velocity to the distance of each galaxy, and we
calculate the genus curve for each sample over a set of smoothing lengths.
We do not enforce non-overlapping volumes for these mock samples,
but the ratio of the simulation volume to the BEST sample volume
is approximately 40:1, so the 27 mock catalogs are approximately
independent.
The error bars in Figure~\ref{gplot} are the standard deviation of
the genus curves from these 27 mock surveys.
The derived statistics 
such as the amplitude ($g$), shift ($\Delta \nu$), cluster abundance ($A_C$), 
and void abundance ($A_V$) parameters are then measured from the genus curves. 

The genus calculated from the observational data suffers from systematic 
effects due to redshift space distortion and complicated survey boundaries. 
To estimate the effects we use the difference between the genus curves 
measured from the real-space galaxy distribution in the whole simulation 
cube and from the redshift-space galaxy distribution in the mock surveys. 
For example, 
we measure the genus density ${g_r}^{\rm cube}$ at a smoothing scale
$R_{G}$ using all halos with mass $M_h > 7.9\times10^{11} h^{-1}M_{\odot}$ 
($\bar{d}=6.1h^{-1}$Mpc)
in the simulation cube.
Halo positions in real space are used for ${g_r}^{\rm cube}$.
The mean genus density ${g_z}^{\rm mock}$ is also calculated from the
 27 mock SDSS surveys
where the halos with $M_h > 7.9\times10^{11}h^{-1}M_{\odot}$ are 
observed in the simulation, with their redshift space positions.
The difference between the two contains the information on the systematic 
effects of survey boundaries and redshift-space distortion (RSD).
The observed genus in real space can be estimated  
by ${g_r}^{\rm obs}={g_z}^{\rm obs}{g_r}^{\rm cube}/{g_z}^{\rm mock}$
assuming that the correction is model-insensitive.
The correction factor is obtained from the ratio of the two statistics 
in the case
of $g, A{_C}$, and $A_V$, and from the difference in the case of $\Delta\nu$.

The correction is expensive because we need to find the correction factors
for all statistics whenever we change the volume-limited sample or the smoothing scale.
But there is a big advantage for presenting the observational data corrected for the survey
boundary effects and RSD effects. 
It enables one to compare the observational data with theoretical predictions 
based on analytic calculations or the analysis of the simulation data without
taking into account the particular survey characteristics.
An empirical study on the systematic effects on the genus is presented 
in Appendix~\ref{app:sysbias}. 

\subsection{Scale-dependent Topology Bias}

\begin{figure}
\epsscale{0.8}
\plotone{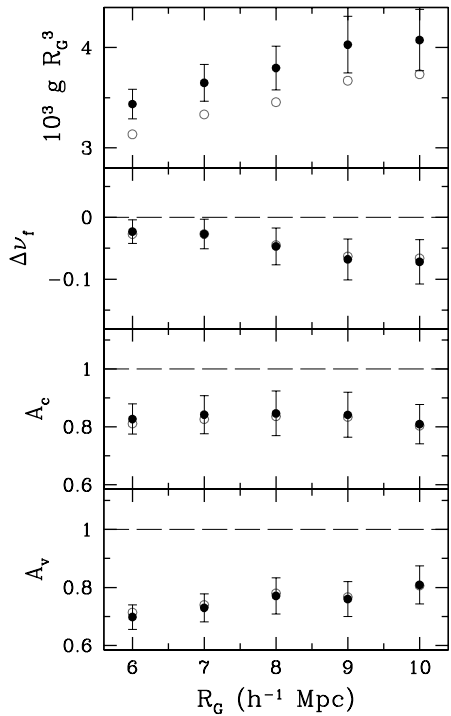}
\caption{Genus-related statistics as a function of Gaussian smoothing
length $R_G$ measured from the BEST sample of galaxies with
$M_r < -20.19$.
In the lower panels, a Gaussian field predicts $\Delta\nu=0$,
$A_C=A_V=1$. Open circles are those before the systematic bias corrections.}
\label{gpara}
\end{figure}

\begin{table*}\tiny
\begin{center}
\caption{Genus-Related Statistics of the Scale-Dependence Samples in Figure~\ref{gpara}}
\label{tab:gparaS}
\begin{tabular}{cccccc}
\hline\hline
\noalign{\smallskip}
Statistics&  \multicolumn{5}{c}{$R_G$}\\
          &      6.0 & 7.0 & 8.0 & 9.0 & 10.0\\
\hline
        $G$&$410.0(373.8)\pm17.9$&$270.3(247.0)\pm13.5$  &$184.7(168.1)\pm 10.6$&$ 133.7(121.8)\pm 9.3$&$  97.5(89.4)\pm 7.3$\\
$\Delta\nu$&$-0.023(-0.027)\pm0.019$&$-0.027(-0.027)\pm0.024$  &$-0.047(-0.045)\pm 0.030$&$-0.068(-0.063)\pm0.033$&$-0.072( -0.066) \pm 0.036$\\
      $A_V$&$ 0.70(0.71)\pm0.04 $&$ 0.73(0.74 )\pm0.05$  &$ 0.77( 0.78)\pm 0.06$&$ 0.76( 0.77)\pm0.06$&$ 0.81(  0.81) \pm 0.07$\\
      $A_C$&$ 0.83(0.81)\pm0.05 $&$ 0.84(0.83 )\pm0.07$  &$ 0.85( 0.84)\pm 0.08$&$ 0.84( 0.83)\pm0.08$&$ 0.81(  0.80) \pm 0.07$\\
\hline
\end{tabular}
\end{center}
{\bf Note.} 
$R_G$ is the smoothing length in units of $\hmpc$.
G is the amplitude of the genus curve,
$\Delta \nu$ is the shift parameter, and
$A_C$ and $A_V$ are cluster and void abundance parameters, 
respectively. All these values are systematic bias-corrected, and 
the observed values before the systematic bias corrections 
are given in parentheses.
Uncertainty limits are estimated from
27 mock surveys in redshift space. 
\end{table*}

To study the scale dependence of 
topology without the effects of changing large-scale structure in
the survey volume, we fix the sample definition 
(i.e. angular and radial boundaries) and change the smoothing length only. 
Figure~\ref{gpara} shows the genus-related statistics 
(filled circles with error bars) as a function of Gaussian smoothing 
length in the case of the BEST sample and the data are given in 
Table~\ref{tab:gparaS}.
Error bars are 
obtained from 27 mock BEST surveys made in the S1 $\Lambda$CDM simulation
using the halos having the mean separation of $6.1 h^{-1}$Mpc. 
Open circles are the statistics before the systematic bias corrections
are made.

Figure~\ref{gpara} demonstrates that the deviation from the Gaussian-field
topology is scale-dependent. 
The second panel shows that genus curve shifts towards increasingly
negative values of $\Delta\nu$ (``meatball'' topology) as
$R_G$ increases from 6 to $10h^{-1}\,$Mpc.
The fractionally largest and statistically strongest deviations
from the Gaussian field prediction arise for the void 
abundance parameter $A_V$, which rises from 0.7 at $6\hmpc$
to 0.8 at $10\hmpc$ (compared to $A_V=1$).
We call these phenomena the scale-dependence of galaxy clustering topology.
The cluster abundance $A_C$ also differs from the Gaussian-field prediction
$A_C=1$, but in a way that is approximately constant with smoothing length 
over this range.
Note that the figures plot 
$gR{_G}^3=(G_{\rm obs}/V_{sample})R{_G}^3$, which is proportional to
the genus per unit smoothing volume.

\begin{figure}
\plotone{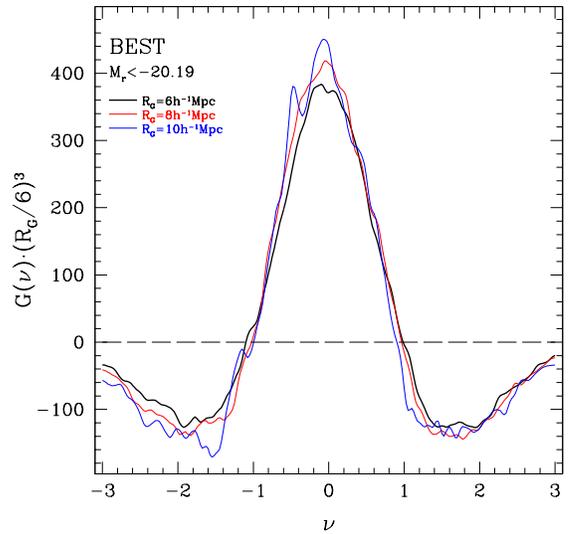}
\caption{Genus curves measured from the BEST sample at three smoothing scales,
$R{_G}=6$ (black line), 8 (red line), and $10\hmpc$ 
(blue line). 
Their amplitudes are scaled by $(R{_G}/6h^{-1}{\rm Mpc})^{3}$.
These curves are not corrected for the systematic biases.}
\label{gplotS}
\end{figure}

To visually relate these measures to genus curve shapes, we 
show the genus curves for $R{_G}=6, 8$, and $10 h^{-1}\,$Mpc,
scaled by $(R{_G}/6h^{-1}{\rm Mpc})^{3}$ in Figure~\ref{gplotS}. 

\begin{figure}
\epsscale{0.8}
\plotone{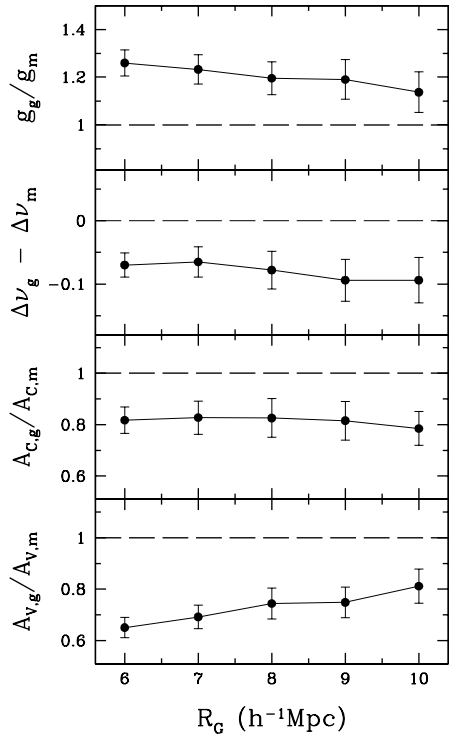}
\caption{Scale-dependence of the galaxy topology bias with respect to 
the matter field. Subscripts $g$ and $m$ represent the SDSS galaxy 
and the matter field, respectively. 
Galaxies in the BEST sample ($M_r < -20.19$) are used to measure the 
galaxy topology. The CDM distribution in our $\Lambda$CDM
simulation is adopted as the matter field.
}
\label{bias}
\end{figure}

To understand the bias in the topology of galaxy clustering we need to know 
the underlying matter field. We assume the matter field 
of the universe is equal to that of our $\Lambda$CDM simulation with WMAP 
3 year parameters. The genus is calculated for the CDM field 
(of the full $1024\hmpc$ simulation cube) as 
in Figures 1 and 2 of Park, Kim \& Gott (2005) and compared to that of 
the observed genus in Figure~\ref{bias} where deviations of 
$g_g / g_m, A_{C,g}/A_{C,m}$, and $A_{V,g}/A_{V,m}$ from 1, or deviation 
of $\Delta\nu_g - \Delta\nu_m$ from 0 indicate the galaxy biasing with respect 
to matter. We find significant bias in the topology 
of the galaxy distribution for all four statistics. 
When compared to the
matter distribution, galaxy clustering shows more complicated structures
(higher $g$), has more meat-ball shifted topology (more negative $\Delta\nu$), 
fewer number of superclusters (smaller $A_C$), and fewer number of voids 
(smaller $A_V$). 
In particular, the void abundance shows a very strong bias which has a strong 
scale-dependence. Any successful galaxy formation model should explain this 
topology bias in the distribution of galaxies. 

James et al. (2007, 2009) studied the scale dependence of topology 
using the 2dFGRS samples.
James et al. (2009) measured the genus statistic 
as a function of scale from 5 to $8\hmpc$, using analytic predictions
for both the linear regime (Gaussian random field) 
and weakly non-linear regime (perturbation theory; Matsubara 1994;
Matsubara \& Suto 1996) to detect the effects of non-linear structure formation
on the genus statistic, and found that both $A_V$ and $A_C$
tend to fall below unity independently of scale and that 
neither of the analytic prescriptions satisfactorily reproduces the measurement.

Quantitative meausures of the shape parameters of the genus curve before 
and after the bias correction shown in Figure~\ref{gpara} help us understand 
the degree of the RSD effects on the genus.
We find that the survey boundary effects are very small
because the volume-to-surface
ratio is quite large for the SDSS DR7 data (see blue solid and black dashed lines
in the top panel of Figure~\ref{sys} in Appendix~\ref{app:sysbias}).

On $6\hmpc$ smoothing scale the shape parameters ($\Delta \nu$, $A_C$,
and $A_V$) change only a little while the amplitude reduction is as much as 
$8.8$\%. 
The RSD effects make the $\nu>1$ part of the genus curve ($A_C$)
have an amplitude lowered by $2.4\%$, and increase the $\nu<-1$ part ($A_V$)
by $1.4\%$ relative to the best-fitting Gaussian genus curve (see Table 3).
Even though the overall amplitude of the genus curve
is rather sensitive to the RSD effects, its shape is not. 
This fact is in agreement with the
perturbative theoretical description of Matsubara \& Suto (1996).

\begin{figure}
\epsscale{0.89}
\plotone{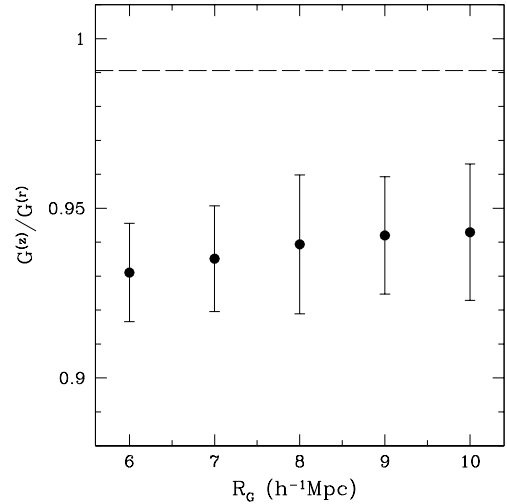}
\caption{Amplitude of the genus curve in redshift space relative to that in real space
from 27 SDSS mock surveys. Dashed line is the linear theory prediction of Matsubara (1996).}
\label{zdist}
\end{figure}

To compare Matsubara (1996)'s linear prediction for the RSD effects with those 
calculated from simulation, we measure the amplitude of the genus curves of 
the 27 mock survey samples both in redshift space and real space, which
are denoted by $G^{z}(\nu)$ and $G^{r}(\nu)$, respectively.
Their ratios are plotted as a function of the smoothing length in Figure~\ref{zdist}.
Dashed line is the linear theory prediction according to Matsubara (1996)
\begin{equation}
\frac {G^{z}(\nu)} {G^{r}(\nu)} = \frac{3\sqrt{3}}{2}\sqrt{u}(1-u),\\
\end{equation}
where 
\begin{equation}
u=\frac{1}{3}\frac{1+(6/5)fb^{-1}+(3/7)(fb^{-1})^{2}}
{1+(2/3)fb^{-1}+(1/5)(fb^{-1})^{2}}.
\end{equation}
The $b$ is the bias parameter and the function $f$ is given by
\begin{equation}
f(\Omega_m,\Omega_\lambda) \equiv d {\rm ln}D/d{\rm ln}a \approx \Omega_m ^{0.6} +
\frac{\Omega_\lambda}{70}(1+\frac{\Omega_m}{2})
\end{equation}
where $D(t)$ is the linear growth factor and $a$ is the scale factor.
The linear theory predicts $G^{z}(\nu)/G^{r}(\nu) = 0.991$, while
our result is 0.931 at $R_G=6\hmpc$. 
$G^{z}(\nu)/G^{r}(\nu)$ tends to slowly approach the linear prediction 
as the smoothing length increases.
Matubara \& Suto (1996) have pointed out
that in weakly non-linear regimes the RSD tends to suppress 
the amplitude more strongly than the linear theory prediction,
in agreement with our result.
Figure 9 also shows that, apart from the sample-to-sample fluctuation
in the observed density field, the sample-to-sample variation in the 
large-scale
peculiar velocity field causes about 1.5\% uncertainty in the RSD correction
on the $R_G=6\hmpc$ scale in the case of the SDSS DR7 sample.
(Note that Fig. 9 shows the effects of RSD only with the density field 
in each sample fixed.)

\subsection{Dependence of Topology on Galaxy Properties}

\begin{figure}
\epsscale{0.7}
\plotone{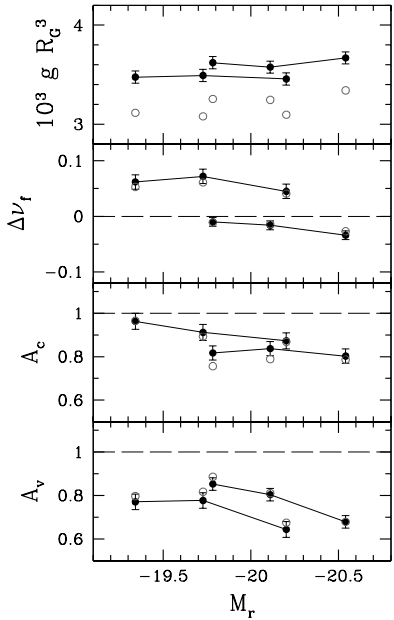}
\caption{Genus-related statistics for subsets of two volume-limited
luminosity samples with $-19.0>M_r>-21.0$ and $-19.5>M_r>-21.5$. 
Three connected points belong to the same volume-limited sample. 
The absolute magnitude ranges of the subsets are  $(-19.5, -20.1), 
(-19.8, -20.5)$, and $(-20.1, -21.5)$ for the brighter sample, and 
$(-19.0, -19.7)$,  $(-19.3, -20.2)$, and $(-19.7, -21.0)$ 
for the fainter sample.
These ranges are chosen so that each subset contains the same number
of galaxies.
All analysis use the same smoothing length of $R_G = 5.9\hmpc$,
and the systematic effects are corrected.
Note that the error bars in this figure do not include cosmic variance ---
they are relevant for comparing the three points in each sample
but not for comparing the samples to each other.
}
\label{gparaL}
\end{figure}

Park et al. (2005) presented the first clear demonstration of
luminosity dependence of galaxy clustering topology:
brighter galaxies show a stronger signal of ``meatball'' topology.
To quantify the luminosity bias we use 
two volume-limited samples with $-19.0>M_r>-21.0$ 
(corresponding to a comoving distance range of
$59.7<R<203.2\hmpc$) and $-19.5> M_r>-21.5$ ($59.7<R<253.0\hmpc$). 
Each sample is divided into three subsets of galaxies 
belonging to three absolute magnitude ranges. 
To be free from confusion due to changing levels of discrete sampling
of structures (random fluctuations in the number of galaxies 
that trace the density field),
we choose the subsamples to contain the same number of galaxies.
Since each of the three subsets includes the same number of galaxies 
distributed in the same 
volume of the universe, any difference in topology must be due to difference 
in luminosity.
We then measure the genus-related statistics to 
quantify the luminosity bias. The statistics measured for $R_G = 5.9h^{-1}$Mpc
are plotted in Figure~\ref{gparaL}. 
The scale is chosen because it is the mean separation of the
galaxies in the subsets of the bright volume-limited sample.
This smoothing length is also used for the subsets of the faint
sample, whose $\bar{d}$ is $5.4\hmpc$.
The $x$-coordinate of each filled circle is the median absolute
magnitude of each subset, and three circles corresponding to three subsets 
of the same volume-limited sample are connected together. 
%

In Figure~\ref{gparaL} there are two sets of connected points corresponding
to the two volume-limited samples.
Differences between the two sets are due to the fact that the samples enclose
the universe with different outer boundaries (i.e., cosmic variance)
and should not be paid attention here. The luminosity dependence of topology
appears within each volume-limited sample. 

To estimate the significance of the luminosity bias we make 20 mock
subsets for each volume-limited sample by selecting $1/3$ of the galaxies
randomly without taking into account luminosity.
Observational and mock subsets have the same sample volume, geometry,
and galaxy number density. They differ only by galaxy luminosity.
We use those mock subsets to estimate the error bars 
in Figure~\ref{gparaL}.
Therefore, they do not contain the cosmic variance, and 
show the significance of luminosity selection only.

It can be seen that
the distribution of brighter galaxies tends to have a stronger ``meatball''
topology signature (more negative $\Delta\nu$) and greater
percolation of underdense regions (lower $A_V$).
But the luminosity dependence for the shift ($\Delta\nu$) parameter 
is somewhat mild, particularly for the faint sample.
We do not detect 
any statistically significant dependence of $g$ on luminosity.
\begin{figure*}
\plotone{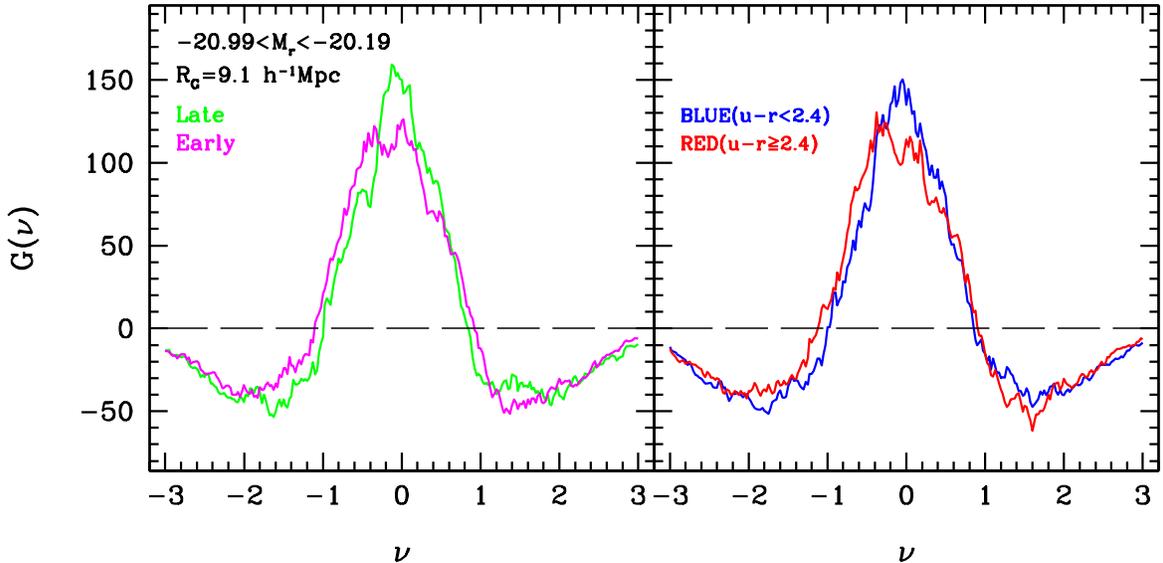}
\caption{Genus curves of early- and late-type galaxies (left panel)
and red ($u-r \ge 2.4$) and blue ($u-r < 2.4$) galaxies 
(right panel) in the BEST sample with narrow luminosity range
($-20.99 < M_r < -20.19$) and at Gaussian smoothing scale of
9.1$h^{-1}$Mpc. These curves are not corrected for the systematic biases.} 
\label{2gplot}
\end{figure*}

After luminosity we consider morphology.
It is well-known that early-type 
galaxies are dominant in massive clusters and late types are abundant 
in the field, and one naturally expects morphology dependence of the topology 
of galaxy clustering. The left panel 
of Figure~\ref{2gplot} shows the genus curves of early- 
(red curve with a smaller amplitude) and late-type galaxies 
with $-20.99 < M_r < -20.19$ at 9.1$h^{-1}$Mpc Gaussian smoothing scale.
We limit the absolute magnitude range to reduce the luminosity-dependent 
bias and to inspect the morphology dependence of the genus only. 

\begin{figure*}
\epsscale{0.7}
\plottwo{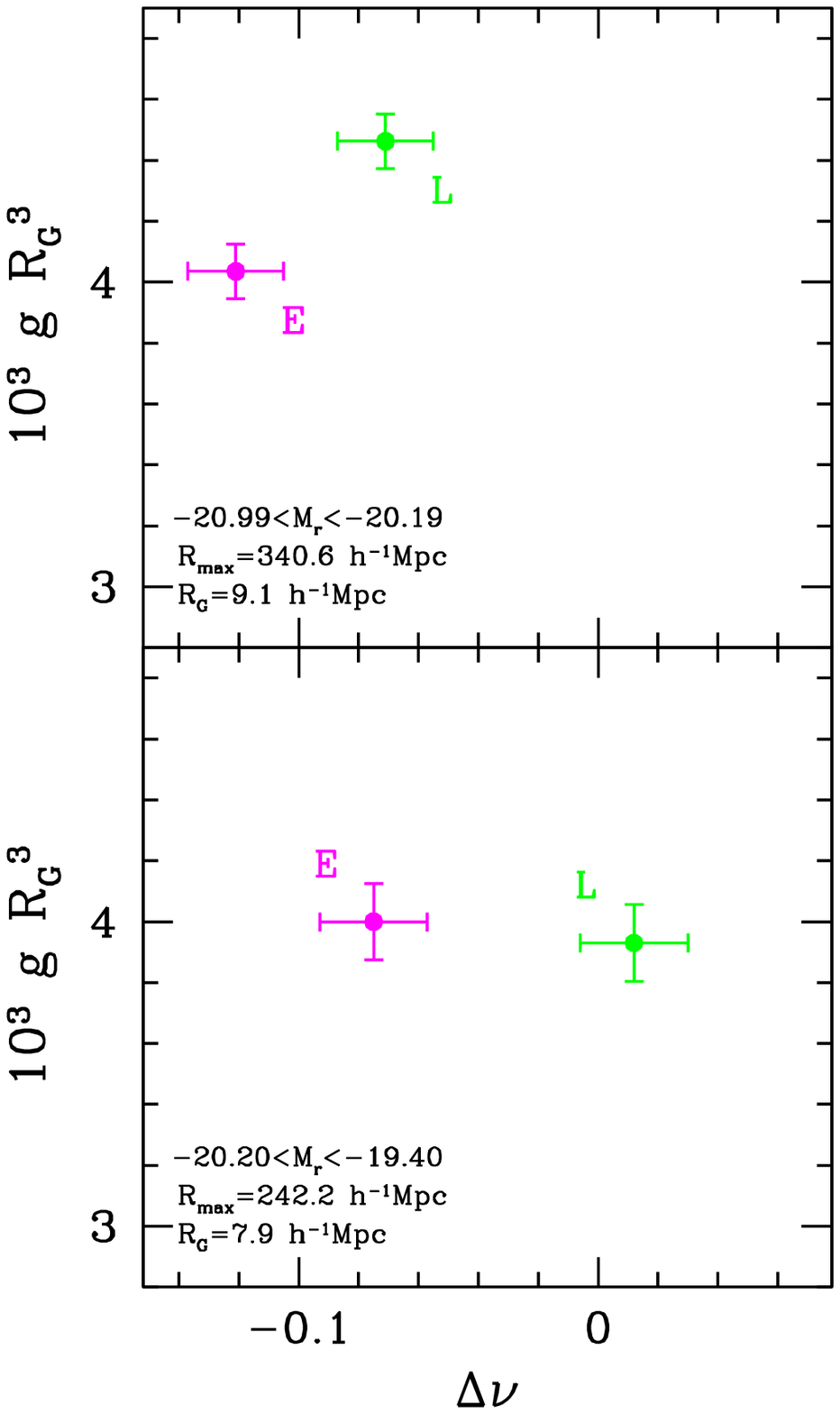}{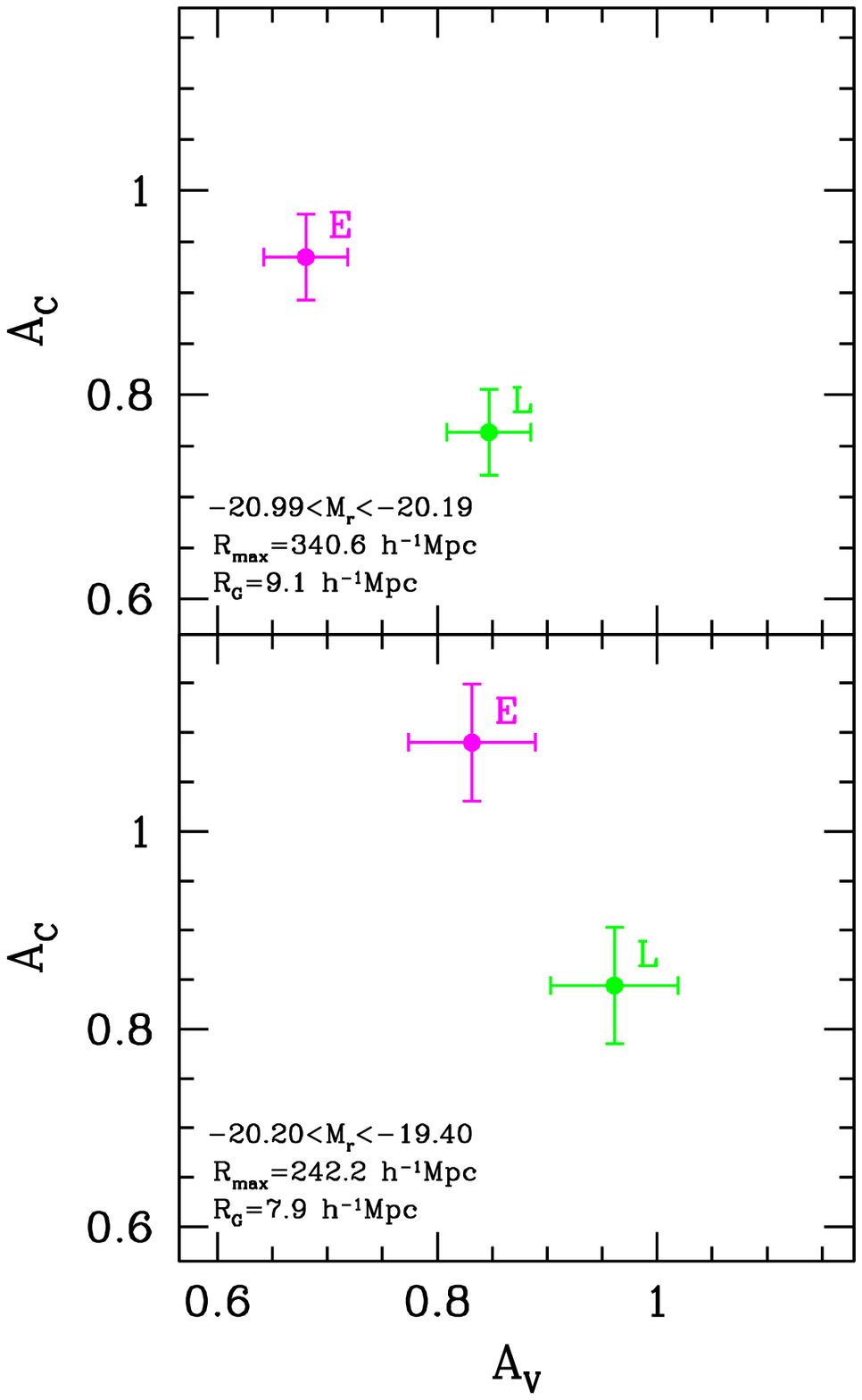}
\caption{Genus-related statistics for the morphology subsamples with narrow
luminosity range. Upper panels are for relatively brighter galaxies with
$-20.99<M_{r} <-20.19$ and the distance limit of
$R_{\rm max}=340.6$h$^{-1}$Mpc. Lower
panels are for fainter galaxies with $-20.20<M_{r}<-19.40$ and
$R_{\rm max}=242.2$h$^{-1}$Mpc.
}
\label{gparaM}
\end{figure*}

We find the mean galaxy separations of the early- and late-type galaxy samples
are 8.9 and 8.3 $h^{-1}$Mpc, respectively.  We discard the most edge-on 
late-type galaxies to make the mean galaxy separation in the late-type sample
equal to 8.9 $h^{-1}$Mpc. 
In Table~\ref{tab:gparaM}, $\bar{d}$ before the thinning out is given 
in parentheses.  The smoothing length of $R_G=9.1 h^{-1}$Mpc is chosen 
because it is the mean separation of 
`blue'-type galaxies, which we will use below to study 
the difference between morphology and color dependence.
The systematic effects are again corrected 
and error bars are estimated from 20 subsets drawn out of the volume-limited
sample made by selecting galaxies randomly without considering morphology.
Each set has the mean galaxy separation of $\bar{d}=8.9 h^{-1}$Mpc.
It can be 
easily seen that the genus curve for early-type galaxies has a smaller overall 
amplitude, fewer voids, and more clusters, and is meat-ball shifted compared to 
that of late-type galaxies. 
These differences are manifest in 
Figure~\ref{gparaM} where we show the genus-related 
parameters for early- and late-type samples separately. 
\begin{table*}\tiny
\begin{center}
\caption{Genus-related Statistics of the Morphology-Dependence Samples in Figure~\ref{gparaM}}
\label{tab:gparaM}
\begin{tabular}{rcccc}
\hline\hline
\noalign{\smallskip}
           &\multicolumn{2}{c}{$-20.19>M_r>-20.99$} & \multicolumn{2}{c}{$-19.40>M_r>-20.20$} \\
           &\multicolumn{2}{c}{$(59.7<R<340.6\hmpc)$} & \multicolumn{2}{c}{($59.7<R<242.2\hmpc$)} \\
           &Early & Late &Early & Late \\
Statistics & $\bar{d}=8.9$ & $\bar{d}=8.9(8.3)$ & $\bar{d}=7.9$ &  $\bar{d}=7.9(6.3)$ \\
\hline
$G$&$128.1 (117.4)\pm 2.9 $&$ 141.7(129.8)\pm  2.9$&$ 66.6( 61.2)\pm 2.1 $ &$65.4 (60.1) \pm 2.1 $\\
$\Delta\nu$&$ -0.121(-0.111)\pm 0.028$&$ -0.071(-0.061)\pm 0.028$&$-0.075(-0.067)\pm 0.018$ &$ 0.012(0.020) \pm 0.018$ \\
$A_V$&$  0.68( 0.67)\pm 0.04$&$  0.85( 0.84)\pm 0.04$&$ 0.83( 0.85)\pm 0.06$ &$ 0.96(0.99) \pm 0.06$\\
$A_C$&$  0.93( 0.91)\pm 0.04$&$  0.76( 0.74)\pm 0.04$&$ 1.09( 1.03)\pm 0.06$ &$ 0.84(0.80) \pm 0.06$\\
\noalign{\smallskip}
\hline
\end{tabular}
\end{center}
{\bf Note.} $\bar{d}$ is the mean galaxy separation in units of $\hmpc$. 
$\bar{d}$ in parentheses is the value before the thinning out 
the late-type sample. 
All genus-related statistics are systematic bias-corrected, and
the observed values before the correction are given in parentheses.
\end{table*}

The upper panels are for the brighter sample mentioned above, and the 
lower panels are for relatively fainter galaxies with $-20.20 < M_r < -19.40$.
In the lower panels, the smoothing length $R_G$ is 7.9 $h^{-1}$Mpc,
which corresponds to the mean separation of the early-type galaxies with
magnitudes in this range (see Table~\ref{gparaM}).
Table~\ref{tab:genusMC} in Appendix~\ref{app:genus} gives
the data for the genus curves.
Table~\ref{tab:gparaM} lists the measured genus-related statistics, 
both corrected and uncorrected (in parentheses) for
systematics.  
It is evident that 
significant morphology bias in the topology of the galaxy distribution exists,
confirming the difference between the early- and late-type distributions
predicted from hydrodynamic cosmological
simulations by Gott, Cen, \& Ostriker (1996).

Next, we examine whether the genus curve depends on galaxy color or not.
Although the morphology and color are strongly correlated
with each other, the roles of morphology and color in galaxy evolution
are expected to be different. For example, galaxy color seems to depend
mainly on the distance and morphology of the nearest neighbor galaxy, 
while morphology depends not only on neighbor but also on the 
distance to the nearest massive cluster of galaxies (Park \& Hwang 2009).
Thus, it is interesting to examine
separately how the galaxy clustering topology depends on
these two physical parameters.

For the color comparison, the genus curves
of red ($u-r \ge 2.4$) and blue ($u-r<2.4$) galaxies with $-20.99< M_r <-20.19$ 
are plotted in the right panel of  Figure~\ref{2gplot}.
We create red and blue samples with the same number 
of galaxies by randomly throwing away some galaxies
in the larger sample. Edge-on galaxies are discarded here.
In Table~\ref{tab:colorcut} of Section 5.3, the mean separation of 
galaxies in each sample and smoothing length are given.
In this color-dependence analysis, we adopted the same sample used for
the morphology-dependence study.
The genus-related statistics for the color subsamples are shown in 
Figure~\ref{gparaMC} together with those of morphology samples for comparison.
Table~\ref{tab:genusMC} in Appendix~\ref{app:genus} gives the data for 
the genus curves, and Table~\ref{tab:gparaC} in Section 5.3 
lists all the measured genus-related parameters.
\begin{figure}
\epsscale{0.75}
\plotone{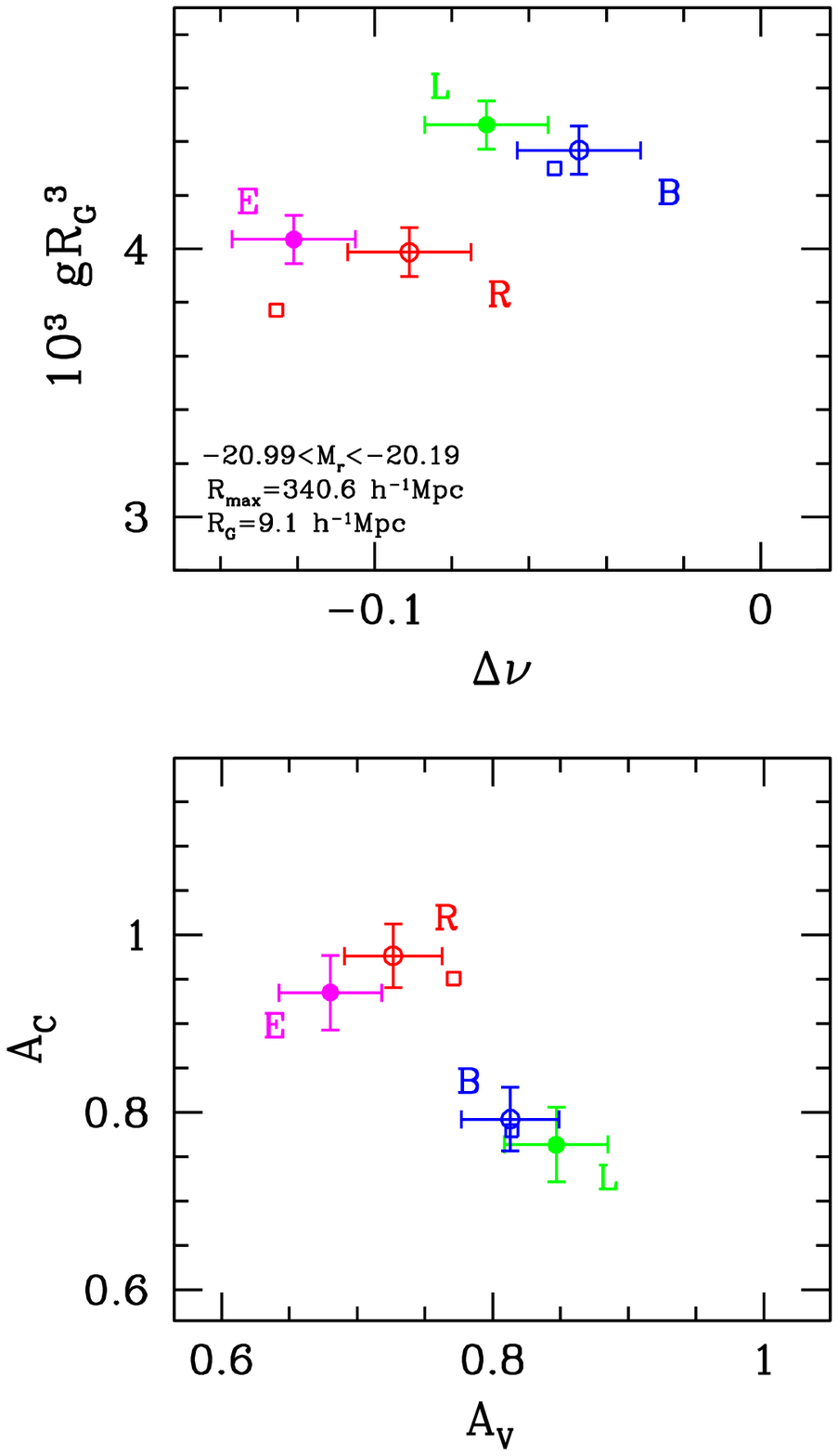}
\caption{Genus-related statistics for the morphology ({\it filled symbols})
and color subsamples ({\it open symbols}) 
with a narrow luminosity range of $-20.99<M_{r} <-20.19$.
Open circles and squares are for color subsets
divided by the color cuts $(u-r)_{\rm cut}=2.4$ and 2.6, respectively. 
}
\label{gparaMC}
\end{figure}
Similarly to the early-type galaxies, the distribution 
of red galaxies has a smaller overall amplitude of the genus curve,
fewer voids, and more clusters, and is meat-ball shifted 
compared to that of blue galaxies. 
The result confirms the relative meatball shift between red and blue galaxies
detected in a 2-D topology survey from the SDSS Early Data Release by 
Hoyle et al. (2002).

There is one major difference between topology of morphology subsets
and color subsets. The contrast in $A_V$ between the early and late 
subsets is much larger than that between red and blue subsets.
This can be also seen for fainter galaxies with $-20.20<M_r<-19.40$
used in Section 5.3.
The dependence of the void abundance on color is much reduced,
while the dependence on morphology still exists in the fainter magnitudes
as shown in the bottom right panel of Figure~\ref{gparaM}.

\section{Test of Galaxy Formation Models}
\subsection{Galaxy Formation Models}
Galaxies are the end product of non-linear gravitational evolution of 
primordial density fluctuations. Therefore, the spatial distribution 
of galaxies and its dependence on the internal properties of galaxies 
should depend both on initial conditions and on galaxy formation and 
evolution processes. This enables us to test galaxy formation models 
as well as the models on primordial fluctuations using topology.
In this section we examine whether or not various models of galaxy formation
are consistent with our measurement of topology of galaxy clustering,
assuming that the difference between observation and models 
is entirely due to inaccuracy in the galaxy formation mechanism.
This assumption is supported by the fact that the topology of the Luminous
Red Galaxies in the SDSS analyzed by Gott et al. (2009) agrees very
well with that of the mock galaxies identified in the $\Lambda$CDM model
with the same cosmological parameters (i.e. 3 year WMAP parameters)
we adopt in this paper.
Gott et al. measured the topology at very large scales of $R_{G}=21$
and $34h^{-1}$Mpc. In this almost linear regime, the results are 
sensitive only to the adopted cosmological parameters.
They use our simple HGC model to locate
the mock LRGs in a large $\Lambda$CDM simulation.
Gott et al.'s results indicate that our cosmological model, which adopts
the initially Gaussian primordial fluctuations with the WMAP 3 year 
$\Lambda$CDM power spectrum, is consistent with the observation in the
linear regime. If one finds a discrepancy between observation and models
in the galaxy clustering topology on non-linear scales, therefore,
it most likely arises from inaccuracy in the galaxy formation models.

We adopt five galaxy allocation 
schemes applied to cosmological N-body simulations. The first is 
the HGC model described in Section 4.2. 
Each of the dark halos identified in our S1 $\Lambda$CDM simulation of 
2048$^3$ CDM particles in a 1024$h^{-1}$Mpc size box is assumed to 
host one galaxy. The halos used in the HGC model can be either isolated
or grouped, and in the latter case they can be the central one or satellites.
The second is a Halo Occupation Distribution (HOD) 
model of Yang et al. (2007) 
refined by introducing a conditional luminosity
function (van den Bosch et al. 2007) that
accurately matches the SDSS luminosity function and the clustering properties
of SDSS galaxies as a function of their luminosity.
The other three models that we examine are different implementations of semi-analytic models
of galaxy formation (SAMs): Croton et al. (2006),  
Bower et al. (2006), and Bertone et al. (2007). 
All of them are set in the context of structure formation
in a CDM universe as modelled by Millennium Simulation (Springel et al. 2005)
of 2160$^3$ particles in a 500$h^{-1}$Mpc size box (
$\Omega_m=0.25, \Omega_{\Lambda}=0.75, \Omega_b = 0.045,
h=0.73, n{_s}=1$ were assumed).
However, they used different construction of the dark matter merger trees
and different implementation of the various physical processes 
involving the baryonic component of the universe.
Croton et al. and Bower et al. invoked `radio-mode' AGN feedback schemes
to restrict gas cooling within relatively massive halos. 
The prescriptions in both models were set by the requirement that
they approximately reproduce the observed break in the present
day luminosity function at bright magnitudes, but their detailed
implementations differ. 
Bertone et al. have used a variation of the SAM of 
de Lucia \& Blaizot (2007), which is the later version of the 
galaxy formation models described in Croton et al., and 
implemented supernova feedback using a `dynamical' treatment of 
galactic winds (Bertone, Stoehr \& White 2005).
The treatment of fast recycling of ejected gas and metals
predicts a relatively low abundance of dwarf galaxies
in better agreement with observations but a rather larger number of 
bright galaxies than observed.

\subsection{Topology Test}
\begin{figure*}
\plottwo{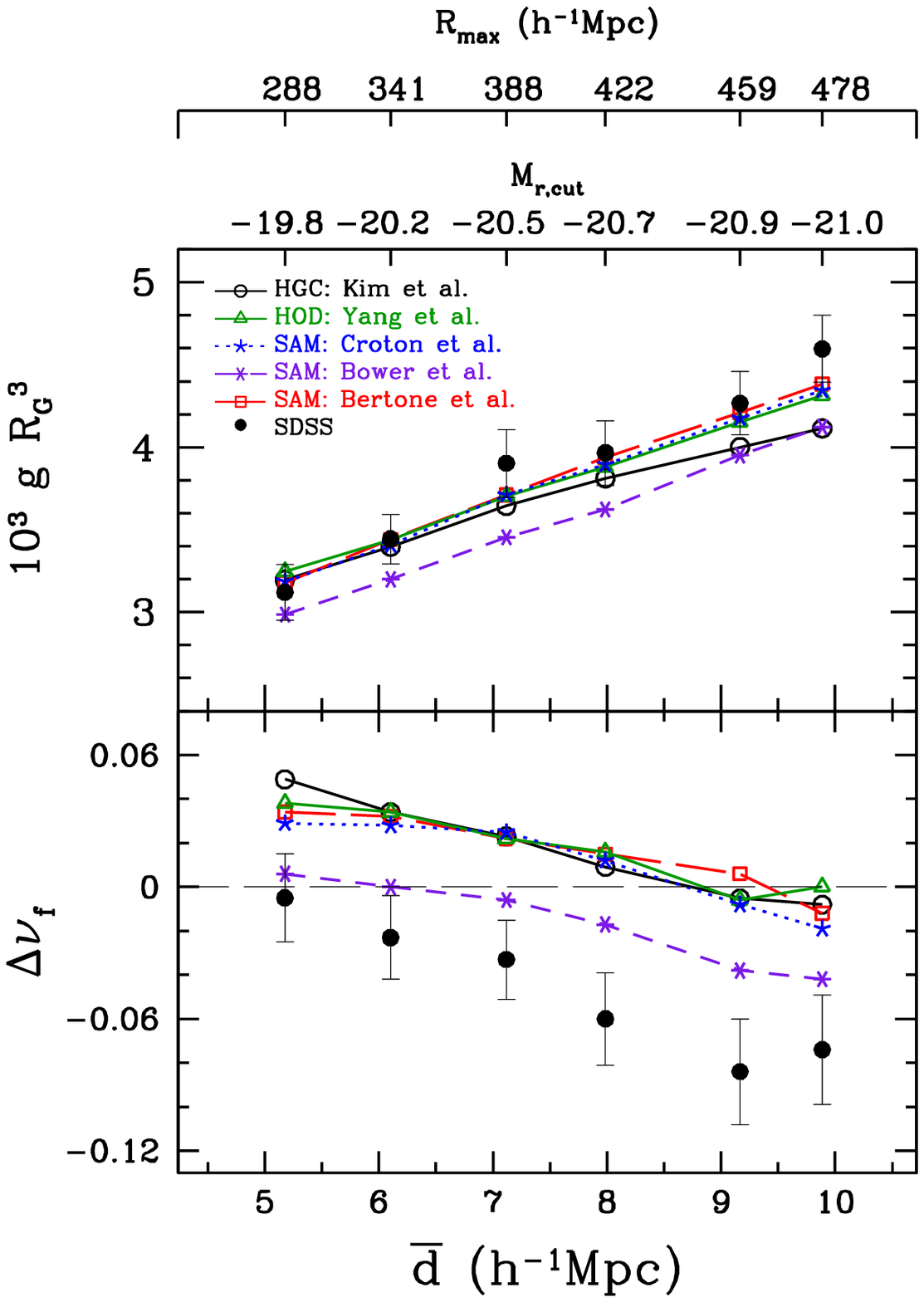}{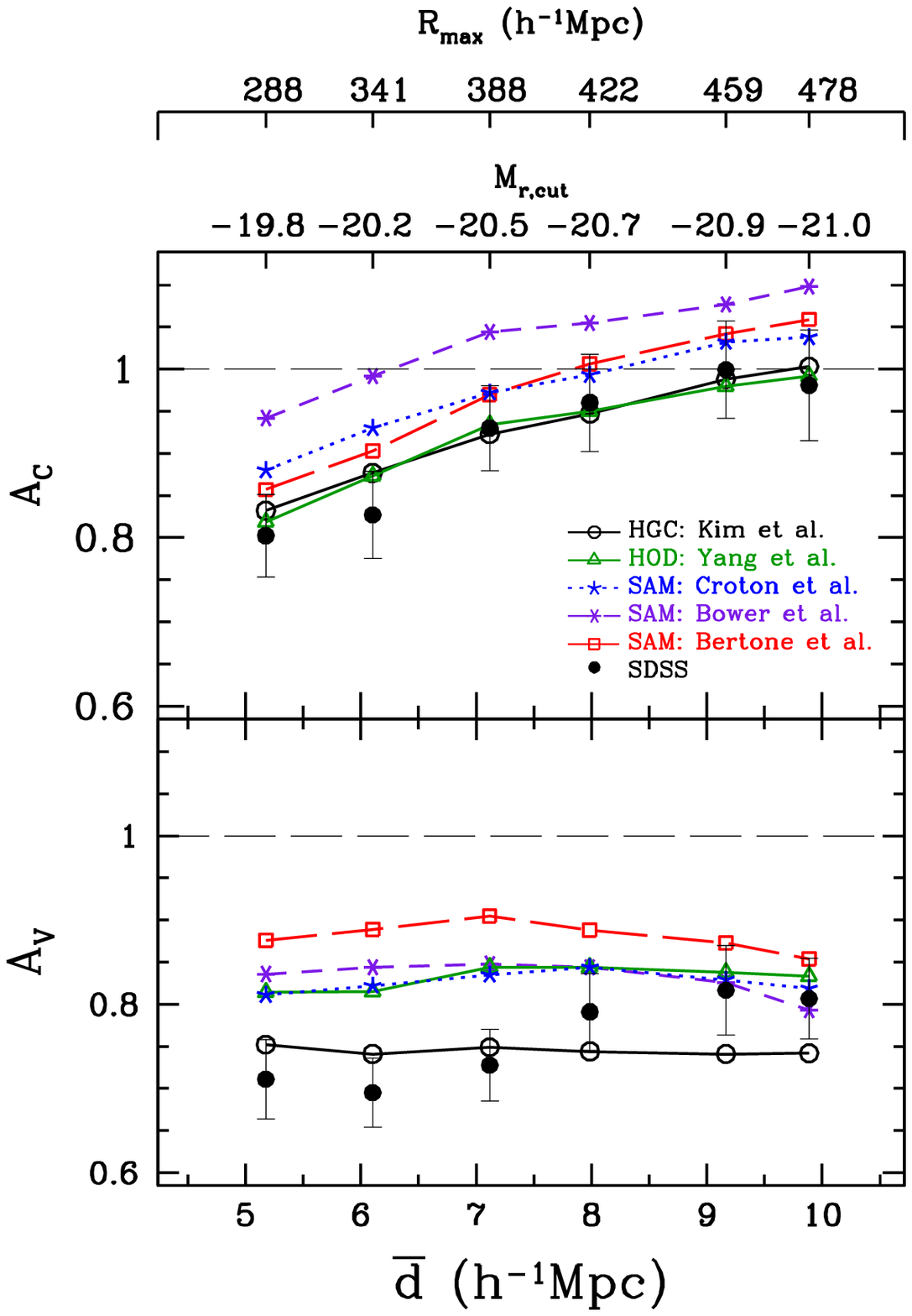}
\caption{Genus-related statistics measured from six volume-limited
samples of SDSS galaxies (filled circles with error bars) 
compared with those of mock galaxies produced by five galaxy formation models.
Results for each model are shown by a line connecting six symbols.
The volume-limited samples are distinguished by the mean galaxy separation 
$\bar{d}$, which also determines the absolute magnitude cut $M_{r,\rm cut}$
or the maximum sample depth $R_{\rm max}$.
Other sample definition parameters $M_{r,\rm cut}$ and $R_{\rm max}$ are
given in the upper $x$-axes.
}
\label{dcut}
\end{figure*}
\begin{figure}
\plotone{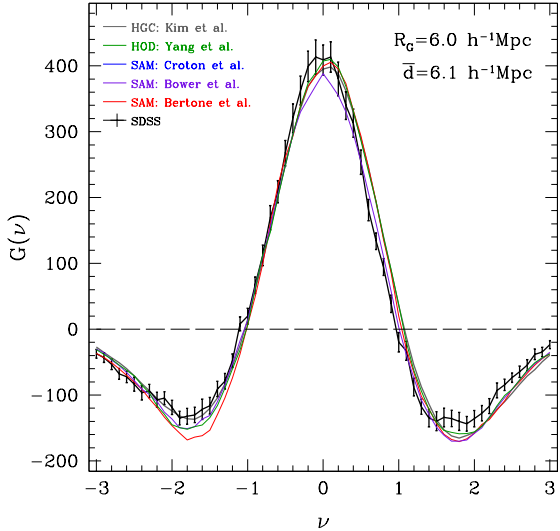}
\caption{Genus curves for the SDSS galaxies (black solid line
with error bars) in the BEST sample and five sets of mock galaxies.
The curve for the observed sample is corrected for the systematic biases.}
\label{gplotGFM}
\end{figure}
\begin{figure}
\plotone{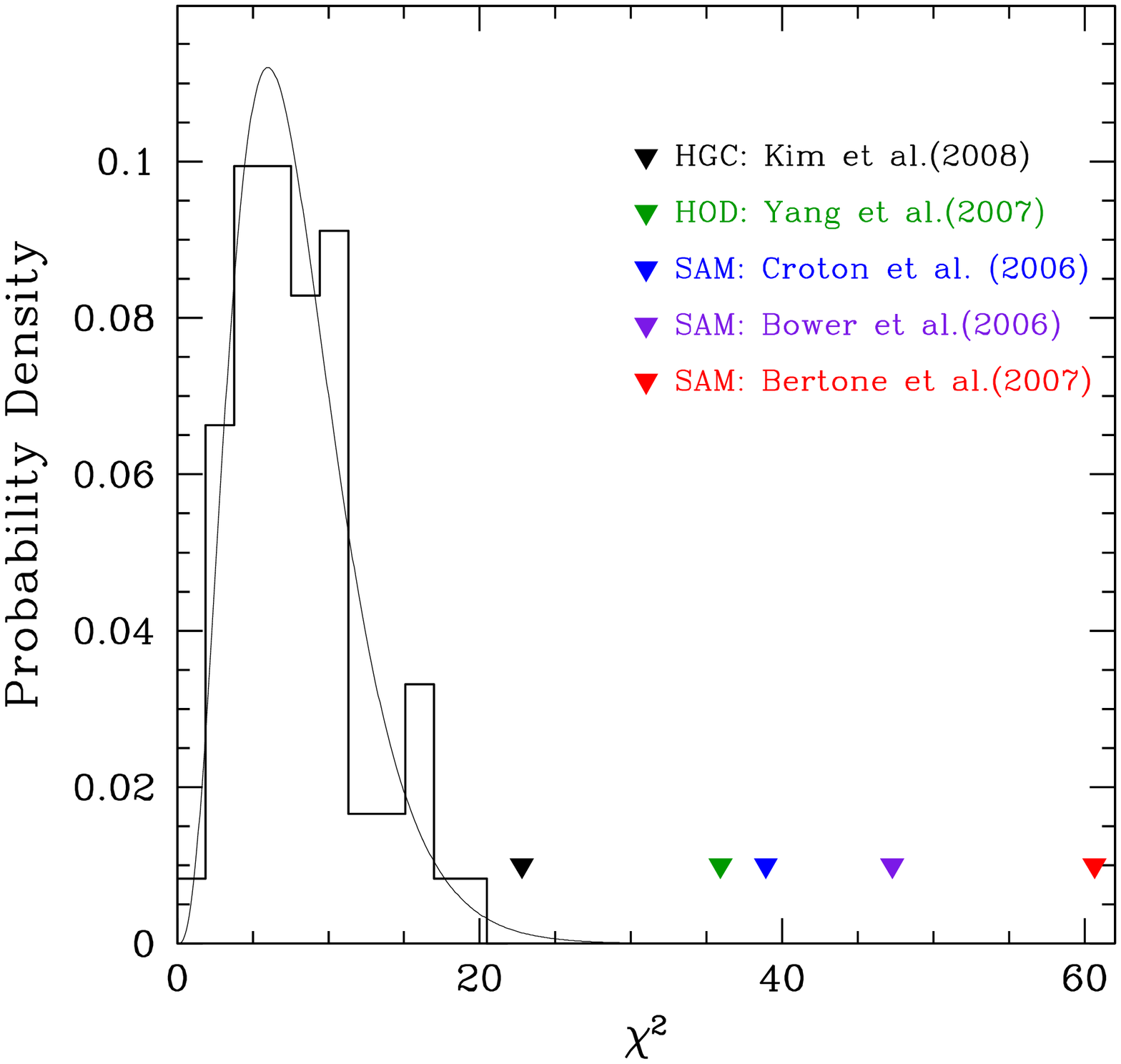}
\caption{Distribution (histogram) of the $\chi^2$ statistic for the mock 
galaxies in the HGC galaxy assignment scheme. The $\chi^2$ statistic  
measures the sample-to-sample fluctuation of the four genus-related 
statistics at two smoothing scales $R_G=6.0$ and $7.2\hmpc$
(see the text for the definition of $\chi^2$).
To obtain the  $\chi^2$ distribution 64 mock SDSS surveys are made 
in the S2 simulation. 
The distribution is fit well by a $\chi^2$ distribution function 
(smooth solid line).
Triangles indicate the $\chi^2$ statistics between the
galaxy models and the observation. 
}
\label{chi2} 
\end{figure}
For a given luminosity cut of a volume-limited sample of the 
SDSS galaxies, the mean galaxy separation $\bar{d}$ and 
the maximum sample depth are determined. 
To compare the mock galaxies generated by the above five galaxy formation 
models in a fair way, we set the mean separation between the mock galaxies 
in each model equal to the observed value
by adjusting the lower limit of galaxy mass or luminosity.
This gives us the relation between $\bar{d}$ and the limiting absolute 
magnitude $M_{r,\rm cut}$. 
The outer boundary $R_{\rm max}$ of the volume-limited sample is 
determined for the apparent magnitude limit of $r=17.6$ and the absolute 
magnitude limit of $M_{r,\rm cut}$.

Comparisons among the observations and models are made for galaxies 
with the same number density or $\bar{d}$. For example, 
we sort all galaxies (halos) in our HGC catalog and find the mass 
cut above which the mean galaxy separation is equal to 
that of each volume-limited sample of the SDSS galaxies. The resulting 
sample of mock galaxies with mass above the cut is to be compared with 
the corresponding observational sample. Mock galaxy subsets are similarly made for the 
HOD and three SAM galaxy samples. The mass or luminosity cuts are given 
in Table~\ref{tab:dcut}.

The genus-related statistics measured from six volume-limited 
samples of the SDSS galaxies (dots with error bars) 
are plotted in Figure~\ref{dcut} at locations 
corresponding to the sample defining parameters 
($\bar{d}$, $M_{r,\rm cut}$, $R_{\rm max}$) which are also given in 
Table~\ref{tab:dcut}.
It should be noted that the observed volume-limited sample contains
progressively brighter galaxies and the sample depth  $R_{\rm max}$ 
increases as the mean galaxy separation  $\bar{d}$ increases.
Figure~\ref{dcut} also shows the results for the mock galaxies of five 
galaxy formation models.  Table~\ref{tab:gparadcut}
lists the statistics for all models and samples.

\begin{table*}
\begin{center}
\caption{Sample Defining Parameters of six volume-limited samples used in Figure~\ref{dcut}}
\label{tab:dcut}
\begin{tabular}{ccccccccc}
\hline\hline
\noalign{\smallskip}
$R_{\rm G}$ & $\bar{d}$ &$R_{\rm min}, R_{\rm max}$&
$M_{r,\rm cut}^{\rm SDSS}   $ &$M_{\rm cut}^{\rm HGC}$&
$M_{r,\rm cut}^{\rm HOD}$&$M_{r,\rm cut}^{\rm Croton}$&
$M_{r,\rm cut}^{\rm Bower }$&$M_{r,\rm cut}^{\rm Bertone}$\\
\hline
  5.2 &5.17&59.7,~288.1&$-19.80$&11.67&$ -19.86$ &$ -20.08$&$ -20.36$&$  -21.25$\\
  6.0 &6.10&59.7,~340.6&$-20.19$&11.90&$ -20.26$ &$ -20.46$&$ -20.72$&$  -21.62$\\
  7.2 &7.11&59.7,~388.3&$-20.50$&12.10&$ -20.56$ &$ -20.75$&$ -21.00$&$  -21.87$\\
  8.0 &7.99&59.7,~422.2&$-20.70$&12.25&$ -20.74$ &$ -20.94$&$ -21.20$&$  -22.03$\\
  9.2 &9.17&59.7,~458.6&$-20.90$&12.43&$ -20.94$ &$ -21.13$&$ -21.42$&$  -22.21$\\
  9.9 &9.89&59.7,~477.8&$-21.00$&12.52&$ -21.04$ &$ -21.23$&$ -21.53$&$  -22.30$\\
\noalign{\smallskip}
\hline
\end{tabular}
\end{center}
{\bf Note.} --
Cols.(1) Gaussian smoothing length in units of $\hmpc$; (2) Galaxy mean separation in units of $\hmpc$;
(3) Inner and outer boundary of the SDSS volume-limited samples in units of $\hmpc$; 
(4) Absolute $r$-band magnitude at $z=0.1$ of the SDSS sample ($K$ and evolution corrected); 
(5) Logarithmic mass cut of halos in HGC model sample in units of $h^{-1}M_{\odot}$;
(6) Absolute $r$-band magnitude at $z=0.1$ ($K$ and evolution corrected) of the HOD galaxy sample;
(7), (8) and (9) Absolute rest frame $r$-band magnitude cuts of three SAM
galaxy samples. 
\end{table*}

\begin{table*}
\begin{center}
\caption{Genus-related Statistics for the samples used in Figure~\ref{dcut}}
\label{tab:gparadcut}
\begin{tabular}{cccccccc}
\hline\hline
Statistics & $R_{\rm G}$ & SDSS &HGC & HOD&
SAM$^{\rm Croton}$ & SAM$^{\rm Bower}$& SAM$^{\rm Bertone}$\\
\hline
$G$
&  5.2 &$343.5(312.7)\pm18.5$ &352.0    &357.6  &350.9  & 328.9 &349.9 \\
&  6.0 &$410.0(373.8)\pm17.9$ &404.6    &409.4  &405.8  & 380.9 &409.8 \\
&  7.2 &$397.2(363.9)\pm20.9$ &371.0    &377.0  &377.8  & 351.6 &377.9 \\
&  8.0 &$377.9(348.9)\pm18.6$ &363.4    &369.6  &371.3  & 345.1 &375.1 \\
&  9.2 &$340.8(317.9)\pm15.3$ &319.4    &331.6  &333.4  & 315.6 &336.2 \\
&  9.9 &$331.7(308.8)\pm14.7$ &297.0    &311.2  &313.5  & 297.7 &316.5 \\
\hline
$\Delta\nu$
&  5.2 &$-0.005(-0.011)\pm0.020$ & 0.049&   0.038&   0.029&  0.006 &   0.034\\
&  6.0 &$-0.023(-0.027)\pm0.019$ & 0.034&   0.034&   0.028&  0.000 &   0.032\\
&  7.2 &$-0.033(-0.030)\pm0.018$ & 0.023&   0.022&   0.025& -0.006 &   0.022\\
&  8.0 &$-0.060(-0.057)\pm0.021$ & 0.009&   0.016&   0.012& -0.017 &   0.015\\
&  9.2 &$-0.084(-0.075)\pm0.024$ &-0.005&  -0.006&  -0.008& -0.038 &   0.006\\
&  9.9 &$-0.074(-0.069)\pm0.025$ &-0.008&   0.000&  -0.019& -0.042 &  -0.012\\
\hline
$A_V$
&  5.2 &$ 0.71(0.73)\pm0.05$ & 0.75  & 0.81 & 0.81  & 0.84&   0.88\\
&  6.0 &$ 0.70(0.71)\pm0.04$ & 0.74  & 0.81 & 0.82  & 0.84&   0.89\\
&  7.2 &$ 0.73(0.75)\pm0.04$ & 0.75  & 0.84 & 0.83  & 0.85&   0.91\\
&  8.0 &$ 0.79(0.81)\pm0.04$ & 0.74  & 0.84 & 0.84  & 0.84&   0.89\\
&  9.2 &$ 0.82(0.84)\pm0.05$ & 0.74  & 0.84 & 0.83  & 0.83&   0.87\\
&  9.9 &$ 0.81(0.82)\pm0.05$ & 0.74  & 0.83 & 0.82  & 0.79&   0.85\\
\hline
$A_C$
&  5.2  &$ 0.80(0.79)\pm0.05$& 0.83& 0.82& 0.88& 0.94&   0.86\\
&  6.0  &$ 0.83(0.81)\pm0.05$& 0.88& 0.87& 0.93& 0.99&   0.90\\
&  7.2  &$ 0.93(0.91)\pm0.05$& 0.92& 0.93& 0.97& 1.04&   0.97\\
&  8.0  &$ 0.96(0.95)\pm0.06$& 0.95& 0.95& 0.99& 1.05&   1.01\\
&  9.2  &$ 1.00(0.98)\pm0.06$& 0.99& 0.98& 1.03& 1.08&   1.04\\
&  9.9  &$ 0.98(0.97)\pm0.07$& 1.00& 0.99& 1.04& 1.10&   1.06\\
\hline
\end{tabular} 
\end{center}  
{\bf Note.} Systematic corrected parameters
Genus-related statistics measured from six volume-limited SDSS samples 
and from mock galaxy samples of five galaxy formation models.  
The smoothing length $R_G$ is in units of $h^{-1}$Mpc.
The observed values are corrected for the systematic biases, and those
before the correction are given in parentheses.
\end{table*}

Figure~\ref{gplotGFM} shows the genus curves for the SDSS galaxies in the BEST sample
and for the five sets of mock galaxies with 6.0 $h^{-1}$ Mpc smoothing.
Table~\ref{tab:genusdcut} in Appendix~\ref{app:genus} gives genus curves of the SDSS and mock galaxies.
Four models,
the exception being Bower et al. (2006), agree very well with one another
for the genus density and shift parameters over the scales explored. 
Those four models agree well with the observations in terms of $g$, 
but Bower et al. (2006) agrees better for $\Delta\nu$. The models give 
diverse results for the cluster and void abundance parameters, though
all of them predict $A_V + A_C < 2$ at all scales shown,
indicating (Matsubara 1994) that effects beyond those
of perturbative non-linear gravitational evolution must be involved.
Bower et al. (2006) is again most discrepant with the observation in terms of 
$A_C$, while HGC and HOD match the observed values almost perfectly.
When $A_V$ is considered, HOD and all SAM models are inconsistent with the
observations, while HGC is within the cosmic variance from the observation.
Overall, no model reproduces all features of the observed topology.
Only the Bower et al. (2006) SAM comes close to matching $\Delta\nu$,
but it fares worst with $A_C$ and the genus density.  Only HGC matches
$A_V$, but it fails for $\Delta\nu$.

To estimate the statistical significance of the failure of each model we
select two volume-limited samples with ${\bar d}=6.10$ and $7.11\hmpc$.
We calculate the $\chi^2$ statistic
$$\chi^2 = \sum_{ij} (v_{ij}^{\rm model} - v_{ij}^{\rm obs})^2 /\sigma_{ij}^2,$$
where the index $i$ runs over the four genus-related
statistics and $j$ runs over the two volume-limited samples. 
$\sigma_{ij}^2$ is the variance of the $i$-th statistic calculated
from 27 mock surveys of the $j$-th volume-limited sample
made in the S1 simulation.
A distribution of this $\chi^2$ statistic is obtained from 64 mock 
volume-limited samples of the HGC galaxies made from the S2 simulation
(see the histogram in Figure~\ref{chi2}),
where $v_{ij}^{\rm obs}$ is replaced by the 
average value over the 64 mock samples.
At each of the 64 locations in the simulation
two mock SDSS samples are made with depths of 340.6 and 388.3 $h^{-1}$Mpc
and with mass cuts of $10^{11.90}$ and $10^{12.10} h^{-1} M_{\odot}$,
respectively.  Smoothing lengths of $R_G=6.0$ and $7.2\hmpc$ are applied 
to these two mock samples, respectively. 

The histogram of $\chi^2$ computed from the mock catalogs in this
way is well described by a $\chi^2$ distribution with 8
degrees of freedom, as expected if the errors in the four statistics
at the two smoothing lengths are essentially uncorrelated.
(Compare the histogram and dotted curve in Figure~\ref{chi2}, and
see further discussion in Appendix~\ref{app:cmatrix_gpara})
Assuming this distribution and applying a 1-tailed $\chi^2$ test,
the probability for the HGC model to be consistent with the observation
is only 0.4\%. The HOD and three SAM models are absolutely ruled out
by this test, confirming that the discrepancies seen in 
Figure~\ref{dcut} are of high statistical significance.
The least inconsistent model is HGC, followed by the HOD model,
while the three SAM models are the most inconsistent.
These discrepancies confirm the findings of Gott et al.\ (2008),
using the DR3 topology measurement and several N-body and hydrodynamic
simulations, that existing galaxy formation models do not reproduce
the topology of the SDSS main galaxy sample near the smoothing scales
we are exploring.
It is interesting to note that the HGC performs best among the five
galaxy formation models even though it is constrained by
only one observational parameter, the mean galaxy number density.

On the other hand, Gott, Choi, \& Park (2009) found
that the topology of the luminous red galaxies (LRGs) is successfully fit 
by the HGC model we have tested here, which means that for the most 
massive galaxies, the HGC model does seem to work well. 
Thus the formation of the most massive LRGs
seems to be a cleaner problem that can be modeled with CDM
simulations and sub-halo finding routines like those described
by Kim \& Park (2006).

There are several implications that can be obtained from this test on
how to fix these galaxy formation prescriptions to make them
reproduce the observed galaxy clustering topology.
The SAM models and HOD model should be modified 
so that the mock galaxies do not separate voids --- connecting
low density regions would lower $A_V$.
The SAM models should also produce more connected
massive clusters (i.e. smaller $A_C$). 
All models should produce large-scale structure
percolating at lower density levels (i.e. more negative $\Delta\nu$).
The problems with the SAM models might be alleviated if the threshold 
interval for galaxy formation were more narrow (i.e., fewer galaxies
in low mass halos), making voids cleaner and superclusters denser.

In principle, of course, the discrepancies in Figure~\ref{gplotGFM}
could be a sign of non-Gaussian initial conditions, or of a linear
power spectrum very different from the one we have adopted based on 
WMAP3.  However, given that different galaxy formation prescriptions
produce differences comparable in magnitude to the observed discrepancies,
even though no one model matches all aspects of the observed
topology, we are inclined to ascribe these discrepancies
to imperfect galaxy formation physics.
The dependence of topology on galaxy morphology and color also
supports this interpretation, as the differences between the topology
of early- and late-type galaxy subsets are larger than the discrepancies
between the full galaxy sample and the observations.

\subsection{Topology Test for Color Subsets}
The SAM models give not only luminosity but also color of mock galaxies,
which makes it possible to compare the observation and SAM models using
the samples divided by color.
We adjust the lower limit of galaxy luminosity
to make the mean galaxy separation $\bar{d}$ equal to that of the SDSS galaxies
in the volume-limited samples with  $-20.2< M_r <-19.4$ (Sample L2) 
and $-20.99 < M_r <-20.19$ (Sample L1). Then each sample is divided
into red and blue subsets using a color cut $(u-r)_{\rm cut}$ that makes
the mean galaxy separation of each color subset equal to the corresponding
color subsets of the SDSS sample.
Table~\ref{tab:colorcut} lists the parameters defining the color subsets for 
the observed and simulated samples.
 
\begin{table*}
\begin{center}
\caption{Sample defining parameters of the color-dependence
samples in Figure~\ref{gparaC}}
\label{tab:colorcut}
\begin{tabular}{ccccccccc}
\hline\hline
\noalign{\smallskip}
     &\multicolumn{4}{c}{Sample L1} & \multicolumn{4}{c}{Sample L2}\\
Samples   &Luminosity & $(u-r)_{\rm cut}$ &$\bar{d}$ &$R_G$&Luminosity& $(u-r)_{\rm cut}$ &$\bar{d}$&$R_G$\\
 \hline
SDSS          &$-20.19>M{_r>}-20.99$ & 2.40 &9.1&9.1&$-19.40>M{_r}>-20.20$ &2.40&6.9 &7.0\\
Croton et al. &$-20.46>M{_r>}-21.21$ & 1.97 &9.1&9.1&$-19.67>M{_r}>-20.44$ &2.18&6.9 &7.0\\
Bower et al.  &$-20.71>M{_r>}-21.51$ & 1.15 &9.1&9.1&$-20.00>M{_r}>-20.68$ &1.20&6.9 &7.0\\
Bertone et al.&$-21.60>M{_r>}-22.28$ & 1.30 &9.1&9.1&$-20.81>M{_r}>-21.59$ &1.77&6.9 &7.0\\
\noalign{\smallskip}
\hline
\end{tabular}
\end{center}
{\bf Note.} Cols. (2) and (5) Absolute $r$-band magnitude limit at $z=0.1$
for the SDSS sample ($K$ and evolution corrected)
and Absolute rest frame $r$-band magnitude limit for the SAM models;
(3) and (7) $u-r$ color cut used for the color subsets;
(4) and (8) mean separation in units of $\hmpc$ between galaxies in the sample;
(5) and (9) smoothing length in units of $\hmpc$.
\end{table*}

\begin{table*}
\begin{center}
\caption{Genus-related statistics for the color-dependence samples in Figure~\ref{gparaC}}
\label{tab:gparaC}
\begin{tabular}{rcccc}
\hline\hline
\noalign{\smallskip}
&\multicolumn{2}{c}{Sample L1} & \multicolumn{2}{c}{Sample L2}\\
 Statistics           &Red & Blue & Red &Blue\\

\hline
& \multicolumn{2}{c}{SDSS samples}&\\
\hline
   $G$ &$126.6(116.2)\pm2.9$ &$ 138.6(127.3)\pm 2.9 $&$  95.6(86.2)\pm 2.7$&$ 101.4(91.4)\pm2.7$\\
$\Delta\nu$
      &$-0.091(-0.091)\pm0.016$ &$ -0.047(-0.047)\pm 0.016$&$-0.056(-0.042)\pm0.016$&$ 0.025(0.039)\pm0.016$\\
$A_V$&$  0.73( 0.73)\pm0.04$ &$  0.81( 0.82)\pm 0.04$&$ 0.82( 0.86)\pm0.05$&$ 0.83(0.87)\pm0.05$\\
$A_C$&$  0.98( 0.93)\pm0.04$ &$  0.79( 0.76)\pm 0.04$&$ 1.01( 0.97)\pm0.04$&$ 0.80(0.77)\pm0.04$\\
\hline
& \multicolumn{2}{c}{Croton et al.}&\\
\hline
$G$        &161.0 &124.7&  99.9&   80.6\\
$\Delta\nu$&0.041  & 0.002& 0.061 & $ -0.050$\\
$A_V$      &0.84  & 0.87& 0.93 & 0.85  \\
$A_C$      &1.06  & 1.04& 0.87 & 1.15  \\
\hline
& \multicolumn{2}{c}{Bower et al.}&\\
\hline
$G$        & 140.5 & 120.7 & 92.4 &  81.2 \\
$\Delta\nu$&  0.004 &$ -0.027$ & $0.044$ & $-0.032$ \\
$A_V$      &  0.94 &  0.85 & 0.92 &  0.85 \\
$A_C$      &  0.96 &  1.12 & 0.84 &  1.16 \\
\hline
& \multicolumn{2}{c}{Bertone et al.}&\\
\hline
$G$        &148.8&  127.6   & 92.4   &84.4   \\
$\Delta\nu$&0.024   &   0.021 &  0.023  & $-0.019$\\
$A_V$      &0.94   & 0.92   & 0.99   & 0.87 \\
$A_C$      &1.02   &  1.01  &   0.82 & 1.07 \\
\noalign{\smallskip}
\hline
\end{tabular}
\end{center}
{\bf Note.} Genus-related statistics for the color subsets of the SDSS samples
and the mock samples of three SAM models.
The observed values before the systematic bias corrections are in parentheses.
\end{table*}
\begin{figure*}
\epsscale{0.79}
\plottwo{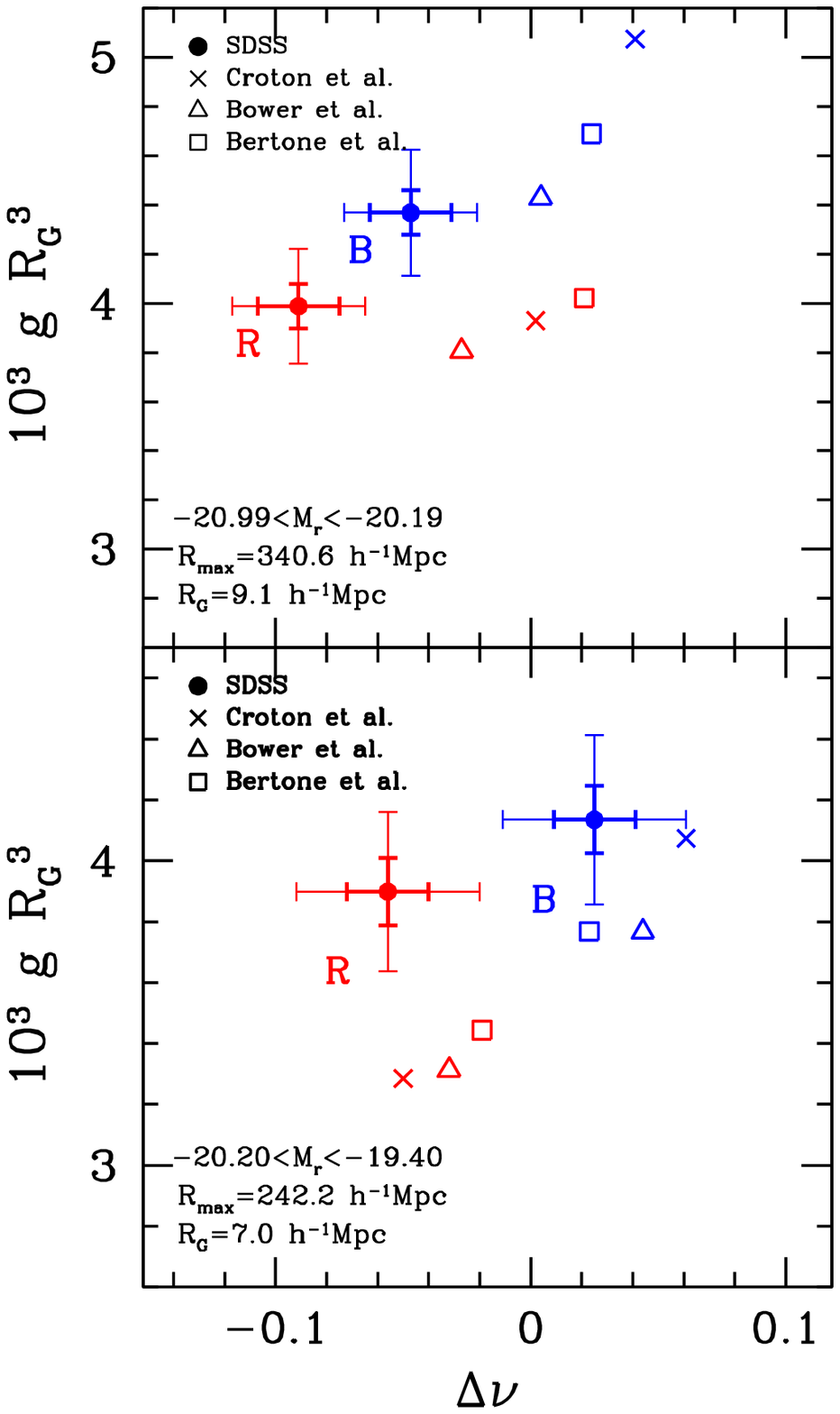}{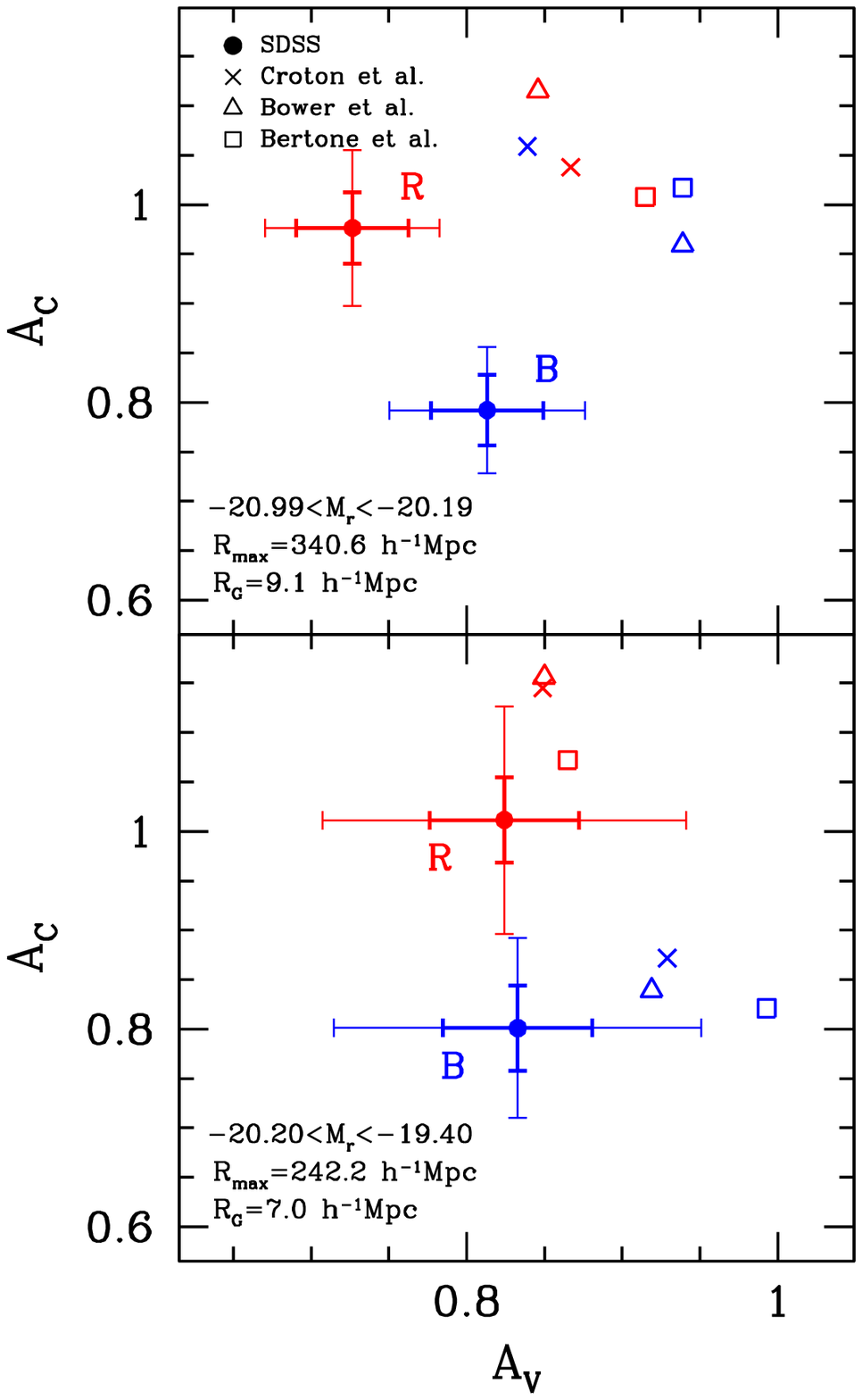}
\caption{Genus-related statistics for the SDSS color subsets
(filled circles) with narrow luminosity ranges.
The upper panels are for the Sample L1 with
$-20.99<M_{r}<-20.19$ and $R_{\rm max}=340.6$h$^{-1}$Mpc, and the lower
panels are for the Sample L2 with $-20.20<M_{r}<-19.40$ and
$R_{\rm max}=242.2$h$^{-1}$Mpc. The thick shorter error bars are for 
significance of color bias. The thin longer error bars
include the cosmic variance, and thus are for comparison between
observation and models. The results obtained by various SAM models 
of galaxy formation are shown by open symbols.
The red and blue symbols correspond to red ($u-r \ge 2.4$ in the case 
of SDSS galaxies) and blue ($u-r<2.4$) galaxy subsets, respectively. 
Square, triangles, and crosses correspond to the SAM models 
of Bertone et al. (2007),
Bower et al. (2006), and Croton et al. (2006), respectively.}
\label{gparaC}
\end{figure*}
Figure~\ref{gparaC} shows the genus-related parameters of the SDSS galaxies
(filled circles with error bars) and mock galaxies (other symbols).
The upper panels are for the relatively brighter sample of galaxies
in L1, and the lower ones are for those in L2.
We show two sets of error bars for the observational data points.
The shorter ones should be used when difference between 
blue and red color subsets is a matter of concern, 
but the longer ones, to which the cosmic variance is included,
should be looked at when a model is compared with the observation.

It can be seen in Figure~\ref{gparaC} that SAM predictions of the 
color-dependence of topology are often qualitatively correct but 
incorrect in magnitude.
Failure of the SAM models in reproducing the observed clustering topology of color
subsets is more evident for the bright blue galaxies of Croton et al.
(2006) in the upper left panel of Figure~\ref{gparaC}.
They have too high genus density and their difference in $g$ with red galaxies
is too large.
All three models predict relatively too positive $\Delta\nu$ for both red
and blue galaxies compared with the combined sample.
The right panels of Figure~\ref{gparaC} show that $A_V$ of both red
and blue mock galaxies is again too large in all three SAM model compared with 
the combined sample. In the case of Bertone et al. and Croton et al., 
the $A_V$ and $A_C$ parameters of the brighter sample (upper right panel) are 
nearly the same for red and blue subsets, which clearly contradicts the 
observations. 

To summarize, the SAM models fail to reproduce the observed clustering
topology of galaxies divided according to color.
Their problems for the combined sample persist in the color subsets, 
and new problems are added. 
In particular, disagreement with the observation is more serious for bright
galaxies.
There should be more bright blue galaxies in superclusters so that
clusters are more connected in the distribution of bright blue galaxies.
Both bright blue and red galaxies should be formed less frequently in
void regions so that voids can be more connected with one another.
The threshold for formation of both red and blue galaxies should be 
at lower density on average
so that the percolation of large-scale structure occurs at lower density.

\section{Summary}
We use the SDSS DR7 main galaxy catalog supplemented with missing redshifts 
and with increased spectroscopic completeness
to measure the galaxy clustering topology over a range of smoothing scales.
The distribution of galaxies observed by the SDSS reveals extremely
diverse structures.

A volume-limited sample, BEST defined by $M_r<-20.19$, enables us to 
measure the genus curve with amplitude of $G=378\pm 18$ at a smoothing scale of
$6h^{-1}$Mpc,
with the quoted uncertainty including all systematics and cosmic
variance.
The amplitude is 5.4 times larger than our previous measurement using
the SDSS DR3 sample (Park et al. 2005) and the uncertainty decreases from 
10.5\% to 4.8\% at the same smoothing length of $R_G=6h^{-1}$Mpc.
We calculate the galaxy clustering topology over the interval from
$R_G=6h^{-1}$Mpc to $10h^{-1}$Mpc, and find mild scale-dependence for 
the shift ($\Delta\nu$) and void abundance ($A_V$) parameters.
The measured genus curve is qualitatively similar to the form
predicted for Gaussian primordial fluctuations (Hamilton et al. 1986),
but the differences are statistically significant at these scales: a shift of
the peak towards negative $\nu$, and fewer isolated voids and
isolated clusters than the Gaussian prediction ($A_C$ and $A_V < 1$).
The bias in topology of galaxy clustering with respect to that of matter is
measured by assuming that the matter density field is given by our $\Lambda$CDM
N-body simulation. We detect strong topology bias in galaxy clustering,
which is also scale-dependent. 

We confirm the luminosity dependence of galaxy clustering topology 
discovered by Park et al. (2005).
The distribution of brighter galaxies is more shifted towards
``meatball' topology (lower $\Delta\nu$) and
shows greater percolation of voids (lower $A_V$).
We find galaxy clustering topology depends also on morphology and color.
Even though early (late)-type galaxies show topology similar to
that of red (blue) galaxies, morphology-dependence of topology is not
identical with color-dependence.
In particular, the void abundance parameter $A_V$ depends on morphology
more strongly than color.

We tested five galaxy formation models, which are used to assign
galaxies to the outputs of N-body simulations.  Three of these
are semi-analytic models, one is an HOD model that assigns galaxies
to halos with parameters tuned to match other clustering statistics,
and one is a scheme that assigns galaxies to halos and subhalos.
None of them reproduces all aspects of the observed topology, though the differences 
from one model to another are comparable to the discrepancies with the observations.  
For this reason, and because the initially Gaussian $\Lambda$CDM model
successfully reproduce the observed topology of LRGs at large scales,
we attribute the discrepancies to failures
of the galaxy formation model rather than non-Gaussian initial conditions.  
The semi-analytic models can also predict the topology
of color subsets, but none of them fully captures the observed
topology differences between red and blue galaxies.

In future work, we will investigate models with non-Gaussian initial
conditions to see what levels of primordial non-Gaussianity can be
ruled out by our measurements.  In principle, the high-precision 
topology measurements presented here and by Gott et al. (2009) can
provide valuable constraints on non-standard inflationary models or
alternative hypotheses for the origin of primordial fluctuations.

Appendix A details our estimates of systematic biases in the genus
curve measurements, demonstrating that the dominant effect is peculiar
velocity distortions in redshift space rather than sample geometry
or boundary effects.  Except where noted otherwise, observational
measurements in this paper are corrected for these biases, so they can
be compared to theoretical predictions in real space with periodic
boundaries.  Appendix B investigates error covariances, showing that
while the individual points on the genus curve have strongly
covariant errors, the statistics $G$, $\Delta\nu$, $A_V$, and $A_C$
are approximately independent.  Appendix C tabulates the full genus
curves for our best samples, complementing the statistics recorded
in earlier tables.

While future surveys will use luminous galaxies and emission-line
galaxies to probe structure in the distant universe, the SDSS DR7
sample is likely to remain the definitive map of large
scale structure at low redshift traced by a broad spectrum of
galaxy types, for the foreseeable future.  The measurements in
this paper characterize the topology of this definitive sample,
attaining unprecedented statistical precision and providing a
valuable test for future models of primordial fluctuations and
galaxy formation physics. 

\acknowledgments
We thank Xiaohu Yang for providing us with the HOD data.
YYC was supported by a grant from the Kyung Hee University in 2010
(KHU-20100179).
CBP acknowledges the support of the National Research Foundation
of Korea (NRF) grant funded by the Korea government MEST (No. 2009-0062868).
JRG is supported by NSF GRANT AST 04-06713.
DHW acknowledges the support of NSF Grant AST-0707985 and an
AMIAS membership at the IAS.
MSV acknowledges the support of NSF GRANT AST-0507647.
SSK was supported by WCU program (R31-1001) and by Basic Science
Research Program (2009-0086824), both through the NRF funded by the MEST 
of Korea.

The Millennium Simulation databases used in this paper
and the web application providing online access to them were
contructed as part of the activities of the German Astrophysical
Virtual Observatory.

Funding for the SDSS and SDSS-II has been provided by the Alfred P. Sloan
Foundation, the Participating Institutions, the National Science
Foundation, the U.S. Department of Energy, the National Aeronautics and
Space Administration, the Japanese Monbukagakusho, the Max Planck 
Society, and the Higher Education Funding Council for England.  
The SDSS Web Site is http://www.sdss.org/.

The SDSS is managed by the Astrophysical Research Consortium for the
Participating Institutions. The Participating Institutions are the
American Museum of Natural History, Astrophysical Institute Potsdam,
University of Basel, Cambridge University, Case Western Reserve University,
University of Chicago, Drexel University, Fermilab, the Institute for
Advanced Study, the Japan Participation Group, Johns Hopkins University,
the Joint Institute for Nuclear Astrophysics, the Kavli Institute for 
Particle Astrophysics and Cosmology, the Korean Scientist Group, the
Chinese Academy of Sciences (LAMOST), Los Alamos National Laboratory,
the Max-Planck-Institute for Astronomy (MPIA), the Max-Planck-Institute
for Astrophysics (MPA), New Mexico State University, Ohio State University,
University of Pittsburgh, University of Portsmouth, Princeton University,
the United States Naval Observatory, and the University of Washington.

This research has made use of the NASA/IPAC Extragalactic Database (NED) 
which is operated by the Jet Propulsion Laboratory, California Institute of Technology,  
under contract with the National Aeronautics and Space Administration.

\clearpage
\appendix
\section{Effects of Survey Systematics on the Genus Curve} \label{app:sysbias}
In this paper we present the observed genus-related statistics corrected
for the systematic effects.
We aim to obtain the statistics in
real space with no boundary and angular selection effects.
Such results can be readily compared with theoretical predictions 
calculated from mock galaxies simulated in a variety of cosmological
models without going through a full analysis taking into account the peculiar
velocities and the complicated angular and radial survey selection functions
of a particular survey.
To understand the effects of the systematics on the genus curve step-by-step,
we first calculate the genus curve from the number density field of the halos
with $M_h >7.9\times10^{11} h^{-1}M_{\odot}$ (${\bar d}=6.1h^{-1}$Mpc) 
in the whole simulation cube (S1 simulation) using periodic boundary conditions.
The halos are the ones identified in our $\Lambda$CDM simulation, which
consist of the isolated halos, the central halos, and the satellite subhalos.
Each of the halos is assumed to contain one galaxy.

\begin{figure}
\epsscale{0.7}
\plotone{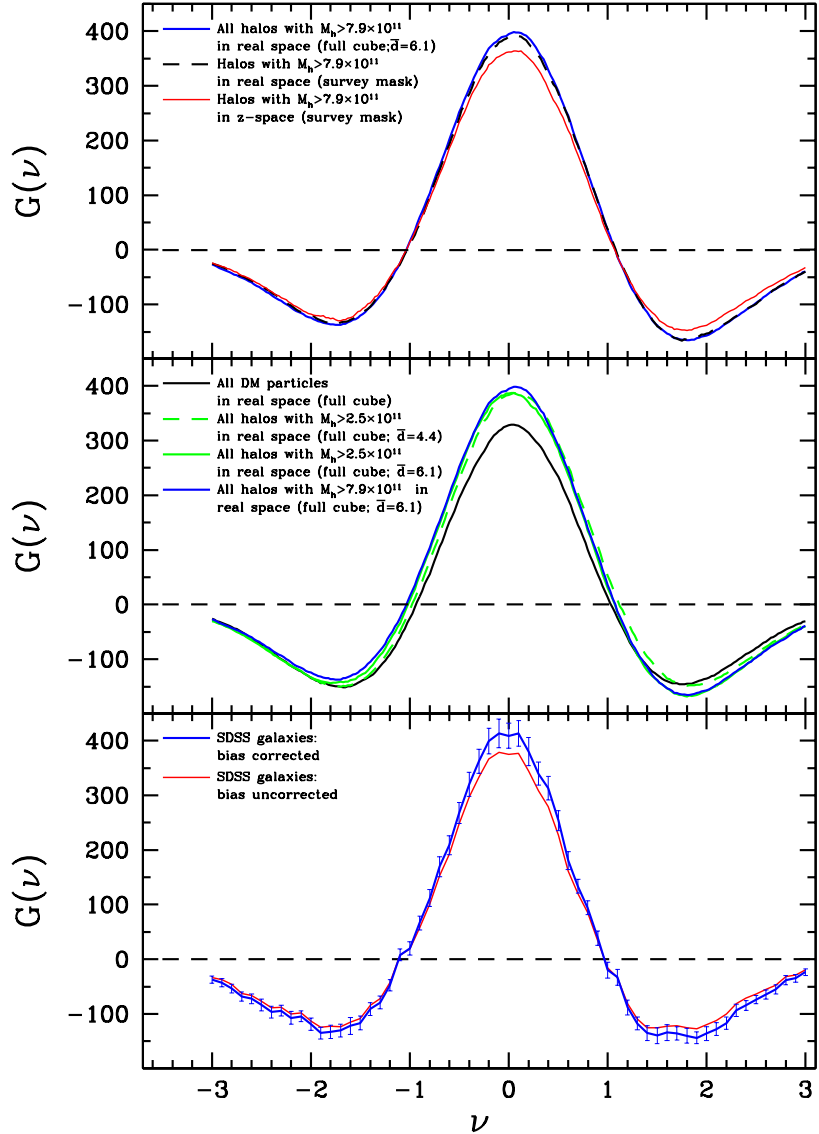}
\caption{
Tests for systematic effects on the genus curve obtained from the HGC galaxies
in the S1 simulation. (top panel) Effects of the SDSS survey mask and 
the redshift space distortion.
The blue solid line is obtained using all HGC galaxies in the whole simulation
cube at their real space positions.
The dashed line and thin solid line are the genus curve averaged over 27 
mock SDSS surveys of the HGC galaxies in real and redshift spaces, respectively.
(middle panel) Effects of different tracers on the genus curve.
The blue solid line is again the real space genus curve from all HGC
galaxies with $M_h >7.9\times10^{11} h^{-1}M_{\odot}$ and 
${\bar d}=6.1h^{-1}$Mpc in the whole simulation cube.
The green solid line is for those with $M_h >2.5\times10^{11} h^{-1}M_{\odot}$
sparsely sampled to have ${\bar d}=6.1h^{-1}$Mpc.
The genus curve of the galaxies  with $M_h >2.5\times10^{11} h^{-1}M_{\odot}$
and with the full sampling (${\bar d}=4.4h^{-1}$Mpc) is shown by 
a green dashed line. The difference shows the effects of shot noise.
The black solid line is the genus curve of the dark matter particle distribution.
(bottom panel) The genus curves of the SDSS DR7 BEST sample before (thin red 
line) and after (blue line with error bars) systematic bias corrections.
A Gaussian smoothing length of $R_{G} = 6.0 h^{-1}$Mpc is used in all cases.}
\label{sys}
\end{figure}

The blue solid line in the top panel of Figure~\ref{sys} is the corresponding 
genus curve when $R{_G}=6h^{-1}$Mpc.
This is the true genus curve we hope to measure with no RSD 
effects and no survey boundary effects (but with some shot noise effects).
We then make 27 mock surveys of these `galaxies' in the simulation cube with radial boundaries
and angular selection function identical to SDSS DR7 trimmed as shown
in Figure 1. The mock galaxies are located at their real-space positions, and 
the resulting mock survey samples are free of the RSD 
effects.
The short-dashed line in the top panel of Figure~\ref{sys} is the genus
curve averaged over these 27 samples.
It can be seen that the radial boundary and angular selection effects
introduce little change in the genus curve in the case of the main part 
of SDSS DR7 thanks to the large volume-to-surface ratio and high
and uniform angular selection function of the sample.
A large bias in the genus curve is generated when galaxies are observed
in redshift space.
The red line in the top panel of Figure~\ref{sys} is the case when both 
RSD effects and the effects due to survey boundaries and angular
selection function are taken into account in the 27 mock SDSS DR7 surveys.
The difference between red line
and dashed line is due to the RSD effects. 
The amplitude of the genus curve is decreased mostly due to the smoothing of structure
by small-scale peculiar velocities of galaxies, but its shape does not change
much.
The genus curve near the median-volume threshold $(\nu=0)$ is slightly pushed
to the negative threshold direction, thus making $\Delta\nu$ decrease.
Both the number of clusters and voids are decreased.
The number of high density regions is decreased more because
massive clusters of galaxies are strongly clustered with one another and two
clusters often appear connected along the line-of-sight in redshift space.
On the other hand, a void can be affected when strong fingers-of-God protrude
into the void from surrounding clusters in redshift space. The number of voids
is decreased only a little between the blue and red lines in 
the top panel of Figure~\ref{sys} because it is difficult for voids 
identified on $R{_G}=6h^{-1}$Mpc scale to be erased by this process.
But the number of voids relative to those expected from the best-fitting
Gaussian curve ($A_V$) is rather increased.

It is also interesting to know how the genus curve depends on the type of
the density tracer.
A comprehensive study on this issue has been made by Park, Kim \& Gott (2005), 
who used a set of N-body simulations of $\Lambda$CDM and the Standard Cold Dark
Matter Universe to examine the dependence of the genus curve on gravitational
evolution, biasing, RSD, smoothing scale, and cosmology.
In the middle panel of Figure~\ref{sys} we show how the genus curve 
changes at $R{_G}=6h^{-1}$Mpc scale 
as the density tracer changes from dark matter particles to 
dark halos with different mass limits.
The genus curve for  the halos with $M_h >7.9 \times 10^{11} h^{-1}M_{\odot}$
($\bar{d}=6.1h^{-1}$Mpc) is plotted by a blue solid line again.
The green solid line is the genus curve for the less massive halos
with mass $M_h >2.5 \times 10^{11} h^{-1}M_{\odot}$
that are sparsely sampled to match the mean galaxy separation to $6.1h^{-1}$Mpc.
The green dashed line is for the full less massive halo data with 
$\bar{d}=4.4h^{-1}$Mpc.
The figure demonstrates the dark matter density field (black solid line)
and the halo number density fields (blue solid line) have very different topology 
at $6h^{-1}$Mpc 
Gaussian smoothing scale.
The $\Delta\nu$ parameter is positive for both matter particles and halo.
The $A_V$ and  $A_C$ parameters for matter are close to 1 and much larger than those of halos.
For the $A_V$ parameter, the difference is much larger.
The strong bias in the observed clustering topology of the SDSS galaxies 
with respect to that of matter is shown in Figure~\ref{bias}. 
On the other hand, the relatively more massive halos
with $M_h >7.9\times10^{11} h^{-1}M_{\odot}$ describe fewer superclusters 
or fewer voids at fixed volume fractions (smaller $A_V$ and  $A_C$) 
when compared with that of halos with $M_h >2.5\times10^{11}h^{-1}M_{\odot}$.

\section{Covariance matrix of the Genus-related statistics}  \label{app:cmatrix_gpara}

\begin{figure}
\epsscale{0.5}
\plotone{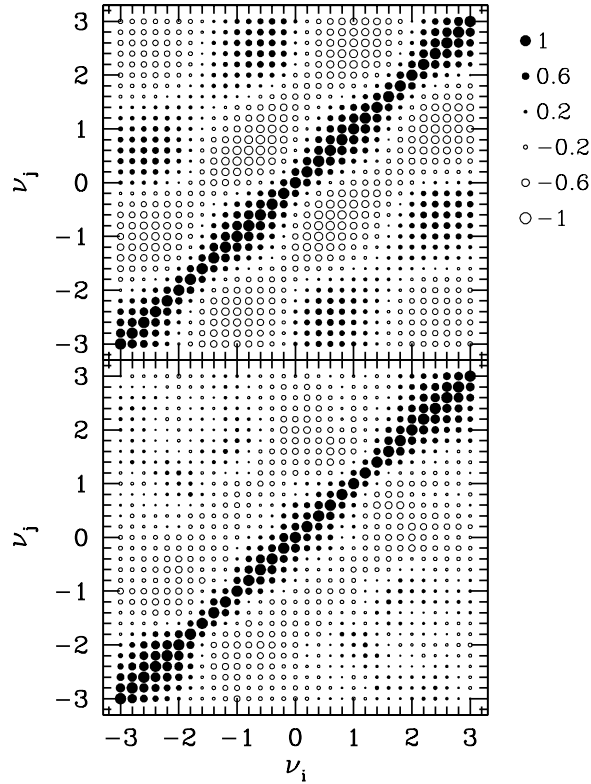}
\caption{Covariance matrix of the genus between two threshold levels 
$\nu_i$ and $\nu_j$. (upper panel) The S1 simulation cube is divided
into 512 subcubes of size $128.0\hmpc$, and  
the number density field of dark matter particles in the subcubes
are used to obtain 512 genus curves and the covariance matrix using $R_G = 6.0\hmpc$.
(lower panel) The S2 simulation cube is divided in 64 subcubes of  
size  $358.4\hmpc$, and the covariance matrix is obtained from
the 64 genus curves of the halo distribution in the subcubes.
Smoothing length of $R_G = 6.0\hmpc$ is used.
Covariance is proportional to the size of circle, and open circles
indicate negative covariances and filled circles indicate positive covariance.}
\label{cov}
\end{figure}

To understand how the measurements of the genus at different levels are 
correlated with one another, in Figure~\ref{cov}
we show the covariance matrix
\begin{equation}
c_{ij}={{<\delta G_{i}\delta G_{j}>}\over{\sigma_{i}\sigma_{j}}}
\end{equation}
where $\delta G_{i} = G(\nu_{i})-\bar{G}(\nu_{i})$ is the difference of
the genus from the mean at a threshold $\nu_{i}$, 
and $\sigma_{i}$ is the standard deviation of the genus at $\nu_{i}$.
Estimation is made using two $\Lambda$CDM simulations, S1 and S2.
(See Table~\ref{tab:sim} for simulation parameters. Detailed explanation of
these simulations is given in section 4.2.)

The bottom panel of  Figure~\ref{cov} shows the covariance measured from 
the halos with $\bar{d}=6.1\hmpc$. The whole simulation cube of the S2
simulation with size of 1433.6$h^{-1}$Mpc is divided into 64 subcubes 
of size $358.4\hmpc$, and the 64 genus curves from these subcubes are used
to calculate the matrix.
The filled and open circles represent positive and negative covariances,
respectively.  The covariance along the diagonal line is one by definition, 
and the radius of circles are proportional to the covariance.
The covariance plot in the upper panel is obtained from the
matter particles with $\bar{d}=0.5\hmpc$ of the S1 simulation
divided into 512 subcubes of size $128.0\hmpc$.  
In both plots a smoothing length of $R_G = 6.0\hmpc$ is used.

The covariance seems to be smallest
for threshold levels $\nu=0$, and $\pm\sqrt{3}$, where the extrema of 
the genus curve are located (see Gott et al. 1990 for a similar finding
for $\nu=0$ level).
There is a positive covariance strip along the diagonal with a width
of about $\sqrt{3}/2$ with the width of positive and negative correlation
ranges slightly depending on $\nu$. 
There are also two negative covariance strips shifted from
the diagonal by $\pm\sqrt{3}$, two positive ones with shifts 
of $\pm 2\sqrt{3}$, and so on. The magnitude of the covariance
decreases rather slowly from the diagonal.

To understand how the genus-related statistics employed in this paper
are correlated with one another, we calculate the covariance matrix
\begin{equation}
d_{ij}={{<\Delta u_{i}\Delta u_{j}>}\over{s_{i}s_{j}}},
\end{equation}
where $u_i$ is the $i$-th one of the four genus-related statistics,
$\Delta u_i=u_i - {\bar u_i}$, and ${\bar u_i}$ is the mean of the $i$-th
statistic calculated from 512 genus curves of the dark matter particles in 
the 512 subcubes of size 128$h^{-1}$Mpc taken from the S1 simulation.
The smoothing length used is $R_G=3h^{-1}$Mpc. The average in Equation (B2)
is taken over 512 subcubes, and $s_i$ is the standard deviation of the $i$-th 
statistic from its mean.  The matrix is symmetric, and its diagonal elements
are one by definition. Figure 19 shows that the genus-related statistics are 
roughly independent of one another, with the magnitude of the covariance
ranging from 0.08 to 0.31.  Thus, it is roughly acceptable
to test a model using a simple $\chi^2$-test with these four statistics
and the quoted observational errors, ignoring covariance, a conclusion
further supported by the mock catalog tests shown in Figure~\ref{chi2}.

\begin{figure}
\epsscale{0.5}
\plotone{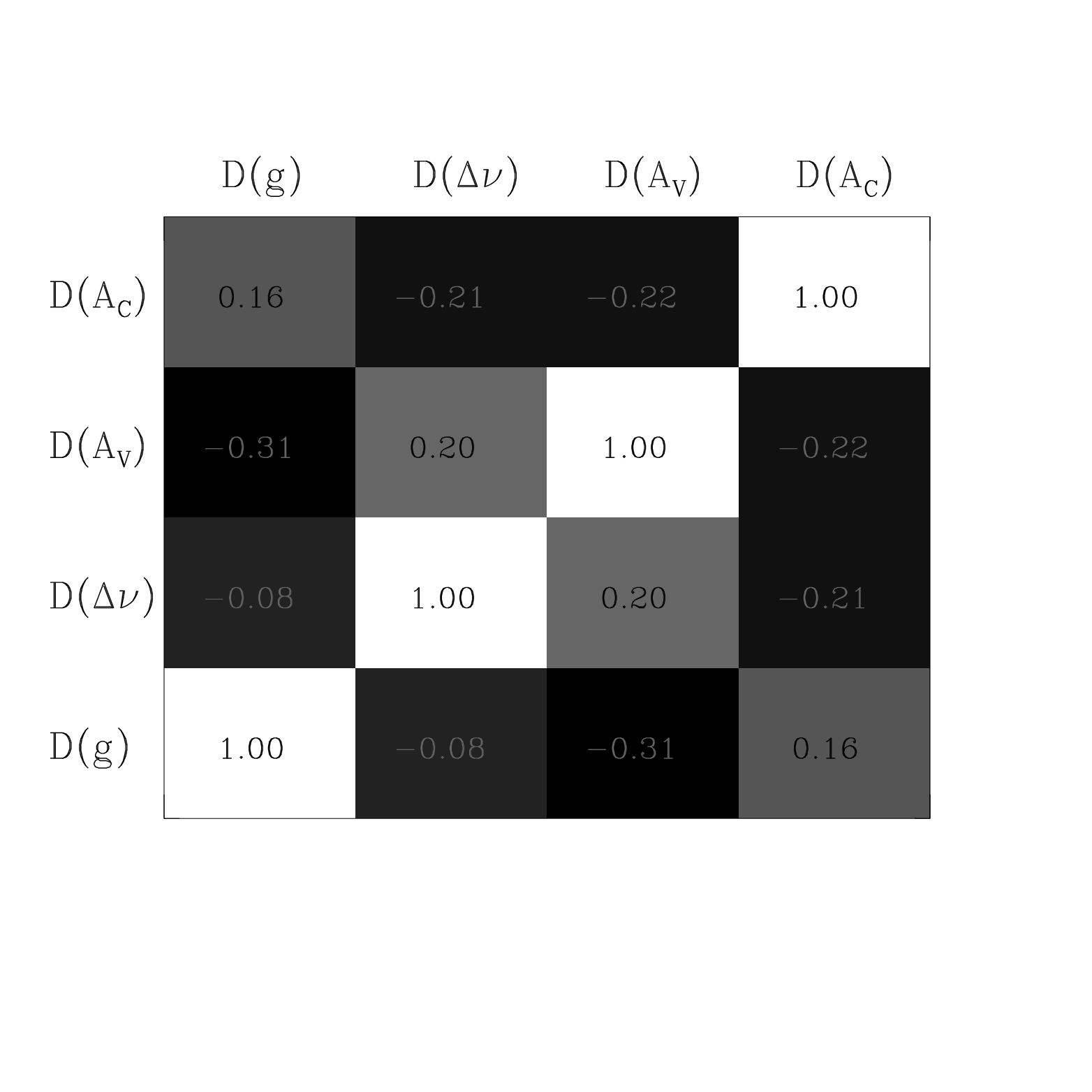}
\caption{Covariance matrix of the genus-related statistics
derived from dark matter particles in 512 subcubes
of the S1 simulation. The matter fields were smoothed with  a
$R_G=3\hmpc$ Gaussian.}
\label{cov_gpara}
\end{figure}

\section{Table of Genus Curves}\label{app:genus}
In this section, we give the tables of the genus curves for the
convenience of readers who wish
to use galaxy clustering topology to test their
galaxy formation or cosmological models.
In Table~\ref{tab:genusMC} 
the genus is given as a function of volume-fraction threshold 
level for the observational morphology and color subsamples.
The genus curves for the SDSS galaxies in the BEST sample and 
the five sets of mock galaxies measured
in Section $5.2$ are given in Table~\ref{tab:genusdcut}.
The genus values of the SDSS galaxies are corrected for the systematic biases,
and the uncorrected values are in parentheses.
These two curves are plotted in the bottom panel of Figure~\ref{sys}. 
Electronic forms of these tables are available from the authors
on request.  Models can also be tested against the genus curve
statistics reported in earlier tables.

\begin{table*}\tiny
\begin{center}
\caption{The Genus values at a given threshold level of the Morphology and
Color Samples in Figure~\ref{gparaM} and \ref{gparaC}}
\label{tab:genusMC}
\begin{tabular}{rrrrrrrrrr}     
\hline\hline
\noalign{\smallskip}
{$\nu$} & \multicolumn{2}{c}{$-20.19<M{_r}<-20.99$}& \multicolumn{2}{c}{$-19.40<M_{r}<-20.40$}&
 \multicolumn{2}{c}{$-20.19<M_{r}<-20.99$} & \multicolumn{2}{c}{$-19.40<M_{r}<-20.40$} \\
        & \multicolumn{2}{c}{($R_G=9.1\hmpc$)}& \multicolumn{2}{c}{($R_G=7.9\hmpc$)}&
 \multicolumn{2}{c}{($R_G=9.1\hmpc$)} & \multicolumn{2}{c}{($R_G=7.0\hmpc$)} \\
      & Early & Late & Early & Late & Red & Blue & Red & Blue \\
\noalign{\smallskip}
\hline
\noalign{\smallskip}
$ -3.0$&$ -14.8$&$ -14.0$&$  -8.8$&$  -7.0$&$ -13.7$&$ -12.6$&$  -9.7$&$  -7.1$\\ 
$ -2.9$&$ -17.8$&$ -17.1$&$  -9.4$&$  -7.2$&$ -20.5$&$ -17.3$&$ -16.5$&$ -10.0$\\ 
$ -2.8$&$ -17.4$&$ -19.6$&$ -12.2$&$ -14.3$&$ -24.2$&$ -21.8$&$ -18.8$&$ -12.4$\\ 
$ -2.7$&$ -21.0$&$ -24.3$&$ -14.8$&$ -15.5$&$ -23.9$&$ -28.6$&$ -15.7$&$ -13.7$\\ 
$ -2.6$&$ -22.8$&$ -28.5$&$ -14.3$&$ -17.8$&$ -28.1$&$ -34.8$&$ -18.4$&$ -18.4$\\ 
$ -2.5$&$ -30.4$&$ -32.8$&$ -14.6$&$ -19.8$&$ -30.2$&$ -35.8$&$ -15.8$&$ -20.2$\\ 
$ -2.4$&$ -29.2$&$ -36.6$&$ -18.7$&$ -20.9$&$ -32.5$&$ -38.9$&$ -22.6$&$ -23.6$\\ 
$ -2.3$&$ -35.4$&$ -43.0$&$ -21.8$&$ -21.9$&$ -44.1$&$ -39.6$&$ -25.1$&$ -26.4$\\ 
$ -2.2$&$ -38.8$&$ -46.4$&$ -21.0$&$ -21.9$&$ -43.1$&$ -42.8$&$ -28.4$&$ -26.2$\\ 
$ -2.1$&$ -42.7$&$ -47.2$&$ -19.0$&$ -26.3$&$ -40.3$&$ -42.4$&$ -32.7$&$ -35.6$\\ 
$ -2.0$&$ -36.7$&$ -47.4$&$ -22.6$&$ -25.9$&$ -38.7$&$ -43.2$&$ -37.4$&$ -35.8$\\ 
$ -1.9$&$ -40.1$&$ -40.1$&$ -18.8$&$ -25.0$&$ -34.4$&$ -48.3$&$ -38.4$&$ -38.0$\\ 
$ -1.8$&$ -39.2$&$ -43.1$&$ -26.1$&$ -32.4$&$ -37.5$&$ -50.4$&$ -34.8$&$ -42.1$\\ 
$ -1.7$&$ -37.1$&$ -54.6$&$ -25.1$&$ -29.1$&$ -40.5$&$ -45.7$&$ -41.4$&$ -44.5$\\ 
$ -1.6$&$ -39.4$&$ -57.9$&$ -23.5$&$ -20.2$&$ -37.7$&$ -40.9$&$ -35.9$&$ -31.1$\\ 
$ -1.5$&$ -26.5$&$ -48.2$&$ -24.7$&$ -21.5$&$ -37.4$&$ -39.7$&$ -29.1$&$ -27.1$\\ 
$ -1.4$&$ -23.9$&$ -44.0$&$ -24.8$&$ -23.1$&$ -36.3$&$ -36.9$&$ -23.5$&$ -29.3$\\ 
$ -1.3$&$ -20.8$&$ -34.9$&$ -20.4$&$ -24.6$&$ -22.6$&$ -43.1$&$ -14.0$&$ -22.9$\\ 
$ -1.2$&$ -16.5$&$ -34.4$&$  -4.4$&$ -11.7$&$ -11.0$&$ -35.6$&$ -13.2$&$ -19.4$\\ 
$ -1.1$&$   3.2$&$ -27.7$&$   4.3$&$  -4.1$&$   4.7$&$ -21.2$&$ -19.7$&$ -10.4$\\ 
$ -1.0$&$  24.3$&$  -5.2$&$   5.7$&$  -1.5$&$  32.1$&$   2.1$&$   1.5$&$  -0.3$\\ 
$ -0.9$&$  55.0$&$  17.9$&$  25.9$&$   1.0$&$  28.4$&$  26.3$&$  31.8$&$  18.6$\\ 
$ -0.8$&$  65.0$&$  49.7$&$  28.3$&$  16.5$&$  46.7$&$  31.6$&$  43.1$&$  21.1$\\ 
$ -0.7$&$  79.5$&$  57.3$&$  41.2$&$  30.4$&$  78.7$&$  45.9$&$  43.6$&$  32.5$\\ 
$ -0.6$&$ 100.7$&$  83.1$&$  49.6$&$  30.6$&$  91.1$&$  69.9$&$  53.5$&$  47.3$\\ 
$ -0.5$&$ 112.1$&$  92.9$&$  37.8$&$  37.3$&$ 105.7$&$  78.6$&$  77.4$&$  51.5$\\ 
$ -0.4$&$ 123.0$&$  82.0$&$  48.3$&$  36.8$&$ 124.3$&$ 108.9$&$  79.9$&$  68.1$\\ 
$ -0.3$&$ 123.6$&$ 121.5$&$  62.0$&$  41.9$&$ 124.0$&$ 137.4$&$  85.5$&$  86.4$\\ 
$ -0.2$&$ 113.3$&$ 151.4$&$  68.9$&$  52.8$&$ 124.7$&$ 146.1$&$  90.4$&$ 100.5$\\ 
$ -0.1$&$ 118.9$&$ 173.3$&$  63.9$&$  65.5$&$ 111.2$&$ 153.0$&$  94.8$&$  98.5$\\ 
$  0.0$&$ 134.5$&$ 161.1$&$  68.6$&$  65.2$&$ 127.0$&$ 150.1$&$  97.6$&$ 102.8$\\ 
$  0.1$&$ 119.0$&$ 157.0$&$  59.2$&$  68.9$&$ 111.5$&$ 143.5$&$  95.5$&$ 110.6$\\ 
$  0.2$&$ 111.7$&$ 118.4$&$  58.1$&$  66.8$&$ 114.3$&$ 128.1$&$  74.5$&$  96.6$\\ 
$  0.3$&$  85.5$&$ 112.1$&$  54.6$&$  68.9$&$  81.4$&$ 100.9$&$  69.6$&$  83.9$\\ 
$  0.4$&$  77.7$&$  96.4$&$  51.0$&$  65.2$&$  81.9$&$ 104.2$&$  78.4$&$  84.9$\\ 
$  0.5$&$  73.8$&$  84.3$&$  40.1$&$  56.7$&$  72.5$&$  85.0$&$  62.3$&$  76.9$\\
$  0.6$&$  52.9$&$  60.8$&$  33.7$&$  35.4$&$  60.1$&$  55.0$&$  48.8$&$  63.1$\\
$  0.7$&$  44.2$&$  32.8$&$  25.7$&$  29.3$&$  49.1$&$  44.2$&$  33.4$&$  34.9$\\
$  0.8$&$  22.1$&$   7.5$&$   9.5$&$   7.9$&$  28.1$&$  15.5$&$  21.1$&$  28.9$\\
$  0.9$&$   6.9$&$ -18.7$&$   4.0$&$  -6.8$&$   1.9$&$ -11.9$&$  12.1$&$  14.5$\\
$  1.0$&$ -12.7$&$ -32.6$&$   0.1$&$ -10.7$&$ -20.6$&$ -20.2$&$  -4.2$&$  -1.0$\\ 
$  1.1$&$ -32.1$&$ -35.0$&$ -14.2$&$ -10.0$&$ -35.5$&$ -22.1$&$ 126.7$&$  90.8$\\  
$  1.2$&$ -44.2$&$ -48.6$&$ -31.4$&$ -15.4$&$ -37.6$&$ -30.4$&$ -36.4$&$ -22.8$\\ 
$  1.3$&$ -57.6$&$ -36.1$&$ -37.1$&$ -26.5$&$ -49.3$&$ -26.8$&$ -40.0$&$ -20.5$\\
$  1.4$&$ -50.6$&$ -30.9$&$ -31.1$&$ -25.1$&$ -45.5$&$ -37.9$&$ -45.4$&$ -31.9$\\
$  1.5$&$ -47.4$&$ -34.5$&$ -30.3$&$ -23.8$&$ -53.5$&$ -45.7$&$ -44.0$&$ -30.5$\\
$  1.6$&$ -46.0$&$ -37.7$&$ -33.5$&$ -25.1$&$ -70.6$&$ -53.9$&$ -40.3$&$ -35.7$\\
$  1.7$&$ -50.7$&$ -44.9$&$ -30.2$&$ -26.7$&$ -56.6$&$ -43.3$&$ -38.1$&$ -38.5$\\
$  1.8$&$ -50.5$&$ -46.1$&$ -23.6$&$ -23.5$&$ -45.1$&$ -42.7$&$ -29.4$&$ -39.7$\\
$  1.9$&$ -40.5$&$ -52.7$&$ -25.9$&$ -20.2$&$ -38.0$&$ -43.5$&$ -33.5$&$ -27.3$\\
$  2.0$&$ -35.9$&$ -46.2$&$ -25.7$&$ -16.4$&$ -36.6$&$ -48.6$&$ -32.0$&$ -29.5$\\
$  2.1$&$ -37.0$&$ -42.0$&$ -20.1$&$ -16.1$&$ -42.7$&$ -42.9$&$ -31.3$&$ -30.4$\\
$  2.2$&$ -38.1$&$ -32.5$&$ -19.7$&$ -14.9$&$ -34.3$&$ -40.0$&$ -32.5$&$ -31.3$\\
$  2.3$&$ -33.6$&$ -35.8$&$ -14.9$&$ -14.3$&$ -33.9$&$ -35.6$&$ -25.6$&$ -21.5$\\
$  2.4$&$ -28.7$&$ -30.7$&$ -15.2$&$ -12.1$&$ -29.7$&$ -29.2$&$ -20.0$&$ -17.6$\\
$  2.5$&$ -24.8$&$ -26.1$&$ -11.5$&$ -11.7$&$ -25.9$&$ -26.4$&$ -16.0$&$ -15.7$\\
$  2.6$&$ -19.4$&$ -22.8$&$ -13.6$&$ -10.0$&$ -18.2$&$ -24.6$&$ -16.9$&$ -15.6$\\
$  2.7$&$ -17.5$&$ -20.3$&$ -11.7$&$  -9.0$&$ -14.1$&$ -21.7$&$ -15.7$&$ -15.2$\\
$  2.8$&$ -12.0$&$ -18.4$&$  -9.8$&$  -6.9$&$ -13.3$&$ -16.6$&$ -11.8$&$ -13.1$\\
$  2.9$&$  -8.6$&$ -11.9$&$  -8.8$&$  -4.3$&$ -11.8$&$ -11.5$&$  -7.4$&$  -9.8$\\
$  3.0$&$  -7.3$&$ -11.0$&$  -5.6$&$  -4.6$&$  -8.2$&$ -10.7$&$  -5.9$&$  -3.1$\\
\noalign{\smallskip}
\hline
 \end{tabular}
\end{center}
\small{\bf Note.} All the values are systematic bias-corrected.
\end{table*}

\begin{table*}\tiny
\begin{center}
\caption{The Genus at a given threshold level of the Samples 
in Figure~\ref{gplotGFM}} 
\label{tab:genusdcut}
\begin{tabular}{r r r r r r r}  
\hline\hline
\noalign{\smallskip}
 $\nu$  &  $G(\nu)_{\rm SDSS}$& $G(\nu)_{\mathrm{HGC}}$ & $G(\nu)_{\mathrm{Bertone}}$ & 
$G(\nu)_{\mathrm{HOD}}$ & $G(\nu)_{\mathrm{Croton}}$ &$G(\nu)_{\mathrm{Bower}}$ \\ 
\noalign{\smallskip}
\hline
\noalign{\smallskip}
%
$ -3.0 $&$  -37.2( -33.6)\pm 6.6$&$   -27.0$&$  -31.0$&$   -30.2$&$   -30.4$&$   -36.8$\\
$ -2.9 $&$  -43.1( -37.0)\pm 7.3$&$   -35.4$&$  -38.0$&$   -37.0$&$   -35.5$&$   -43.0$\\
$ -2.8 $&$  -53.3( -45.5)\pm 7.9$&$   -43.1$&$  -45.4$&$   -43.8$&$   -46.9$&$   -51.0$\\
$ -2.7 $&$  -67.9( -62.5)\pm 8.2$&$   -50.9$&$  -54.3$&$   -52.8$&$   -55.1$&$   -60.2$\\
$ -2.6 $&$  -72.2( -65.8)\pm 8.8$&$   -60.0$&$  -62.3$&$   -61.9$&$   -65.6$&$   -70.1$\\
$ -2.5 $&$  -83.2( -76.1)\pm10.8$&$   -70.8$&$  -70.5$&$   -72.2$&$   -77.9$&$   -80.2$\\
$ -2.4 $&$  -96.5( -88.8)\pm10.8$&$   -82.6$&$  -84.7$&$   -82.9$&$   -90.9$&$   -90.9$\\
$ -2.3 $&$  -93.9( -88.1)\pm10.1$&$   -93.4$&$  -97.1$&$   -95.8$&$   -98.9$&$  -105.7$\\
$ -2.2 $&$ -107.0( -99.5)\pm11.3$&$  -105.4$&$ -116.2$&$  -109.8$&$  -109.6$&$  -118.7$\\
$ -2.1 $&$ -104.5( -98.2)\pm 9.8$&$  -116.1$&$ -129.2$&$  -125.7$&$  -125.9$&$  -133.5$\\
$ -2.0 $&$ -118.6(-112.7)\pm10.8$&$  -124.1$&$ -145.2$&$  -141.1$&$  -136.6$&$  -146.5$\\
$ -1.9 $&$ -135.0(-124.0)\pm10.5$&$  -131.6$&$ -149.5$&$  -149.5$&$  -149.7$&$  -157.8$\\
$ -1.8 $&$ -132.2(-122.1)\pm12.1$&$  -136.0$&$ -152.0$&$  -153.9$&$  -151.0$&$  -168.1$\\
$ -1.7 $&$ -130.1(-123.7)\pm11.6$&$  -136.6$&$ -148.1$&$  -151.8$&$  -148.1$&$  -163.9$\\
$ -1.6 $&$ -121.6(-115.2)\pm15.5$&$  -131.0$&$ -145.4$&$  -145.0$&$  -136.4$&$  -161.7$\\
$ -1.5 $&$ -116.3(-108.1)\pm12.5$&$  -120.0$&$ -134.3$&$  -132.1$&$  -131.0$&$  -153.4$\\
$ -1.4 $&$  -90.8( -84.3)\pm11.7$&$  -104.4$&$ -115.8$&$  -112.1$&$  -110.0$&$  -129.4$\\
$ -1.3 $&$  -78.9( -72.2)\pm11.9$&$   -83.3$&$  -89.7$&$   -96.5$&$   -84.3$&$  -104.7$\\
$ -1.2 $&$  -47.7( -43.1)\pm10.3$&$   -55.5$&$  -60.0$&$   -69.3$&$   -52.2$&$   -74.0$\\
$ -1.1 $&$    8.0(   7.6)\pm10.6$&$   -24.5$&$  -31.4$&$   -31.0$&$   -17.2$&$   -37.6$\\
$ -1.0 $&$   19.7(  19.3)\pm12.1$&$    14.8$&$    6.1$&$    10.0$&$    13.3$&$     6.9$\\
$ -0.9 $&$   66.7(  57.2)\pm12.6$&$    59.4$&$   60.0$&$    58.6$&$    63.3$&$    48.7$\\
$ -0.8 $&$  112.0( 101.1)\pm15.7$&$   102.9$&$  112.1$&$   105.1$&$   109.2$&$    98.5$\\
$ -0.7 $&$  169.4( 153.2)\pm18.3$&$   152.2$&$  147.1$&$   156.7$&$   155.3$&$   159.8$\\
$ -0.6 $&$  208.7( 192.5)\pm17.1$&$   201.1$&$  200.4$&$   205.1$&$   204.9$&$   218.3$\\
$ -0.5 $&$  267.3( 247.8)\pm19.3$&$   250.7$&$  247.7$&$   254.3$&$   247.7$&$   260.9$\\
$ -0.4 $&$  320.3( 296.7)\pm21.9$&$   296.0$&$  295.8$&$   294.4$&$   295.6$&$   293.0$\\
$ -0.3 $&$  362.4( 334.0)\pm22.0$&$   335.2$&$  338.6$&$   340.9$&$   329.8$&$   339.5$\\
$ -0.2 $&$  399.1( 366.7)\pm23.3$&$   368.5$&$  373.8$&$   369.3$&$   350.2$&$   369.1$\\
$ -0.1 $&$  413.3( 379.0)\pm25.9$&$   385.9$&$  389.7$&$   383.1$&$   369.5$&$   384.9$\\
$  0.0 $&$  408.9( 374.6)\pm23.0$&$   395.0$&$  407.4$&$   403.0$&$   388.0$&$   399.5$\\
$  0.1 $&$  413.4( 376.7)\pm23.1$&$   398.3$&$  409.6$&$   402.6$&$   372.2$&$   405.5$\\
$  0.2 $&$  379.9( 344.0)\pm25.8$&$   387.7$&$  398.3$&$   385.3$&$   354.9$&$   394.8$\\
$  0.3 $&$  339.6( 308.6)\pm21.1$&$   362.0$&$  368.7$&$   361.9$&$   329.8$&$   373.0$\\
$  0.4 $&$  312.7( 279.8)\pm21.5$&$   329.4$&$  331.6$&$   326.5$&$   296.9$&$   332.7$\\
$  0.5 $&$  253.7( 226.8)\pm18.4$&$   289.4$&$  286.8$&$   295.2$&$   254.7$&$   293.4$\\
$  0.6 $&$  180.8( 161.2)\pm16.3$&$   245.1$&$  239.0$&$   250.0$&$   206.3$&$   245.0$\\
$  0.7 $&$  133.5( 120.6)\pm12.6$&$   191.8$&$  193.4$&$   186.2$&$   159.8$&$   194.0$\\
$  0.8 $&$   94.3(  85.6)\pm12.8$&$   140.3$&$  139.9$&$   132.3$&$   110.2$&$   135.8$\\
$  0.9 $&$   38.5(  33.0)\pm13.4$&$    89.0$&$   90.1$&$    74.6$&$    57.4$&$    85.5$\\
$  1.0 $&$  -19.7( -15.6)\pm14.8$&$    35.8$&$   37.4$&$    32.5$&$     3.0$&$    19.9$\\
$  1.1 $&$  -32.6( -33.8)\pm14.4$&$   -11.9$&$  -19.9$&$   -26.5$&$   -39.5$&$   -29.8$\\
$  1.2 $&$  -87.8( -80.7)\pm13.6$&$   -53.9$&$  -61.9$&$   -68.7$&$   -73.4$&$   -65.4$\\
$  1.3 $&$ -117.6(-108.8)\pm11.6$&$   -87.4$&$  -93.8$&$  -103.0$&$  -107.6$&$  -101.6$\\
$  1.4 $&$ -135.0(-124.8)\pm13.3$&$  -113.9$&$ -121.6$&$  -127.5$&$  -129.6$&$  -121.1$\\
$  1.5 $&$ -139.7(-125.3)\pm14.3$&$  -137.4$&$ -141.7$&$  -146.0$&$  -148.9$&$  -143.2$\\
$  1.6 $&$ -134.0(-122.1)\pm14.3$&$  -152.9$&$ -153.7$&$  -163.7$&$  -160.9$&$  -159.4$\\
$  1.7 $&$ -135.6(-121.3)\pm12.4$&$  -162.8$&$ -156.5$&$  -172.8$&$  -170.1$&$  -167.4$\\
$  1.8 $&$ -140.0(-124.8)\pm13.6$&$  -165.2$&$ -158.8$&$  -168.3$&$  -169.9$&$  -170.7$\\
$  1.9 $&$ -143.5(-127.3)\pm11.1$&$  -161.0$&$ -158.6$&$  -169.3$&$  -168.3$&$  -166.4$\\
$  2.0 $&$ -135.7(-119.5)\pm12.5$&$  -156.8$&$ -155.7$&$  -160.2$&$  -158.2$&$  -158.0$\\
$  2.1 $&$ -127.6(-111.7)\pm11.8$&$  -146.5$&$ -147.7$&$  -149.1$&$  -149.7$&$  -144.8$\\
$  2.2 $&$ -116.8(-100.0)\pm11.3$&$  -134.4$&$ -133.7$&$  -141.9$&$  -132.7$&$  -137.0$\\
$  2.3 $&$  -93.0( -81.1)\pm11.3$&$  -121.5$&$ -116.0$&$  -129.4$&$  -118.3$&$  -122.8$\\
$  2.4 $&$  -84.4( -71.3)\pm 8.6$&$  -110.9$&$ -100.8$&$  -114.4$&$  -104.1$&$  -108.8$\\
$  2.5 $&$  -73.7( -63.6)\pm 7.2$&$   -97.4$&$  -91.1$&$   -96.9$&$   -91.7$&$   -91.3$\\
$  2.6 $&$  -64.4( -54.7)\pm 8.6$&$   -83.6$&$  -77.1$&$   -81.0$&$   -77.3$&$   -79.8$\\
$  2.7 $&$  -54.6( -47.2)\pm 8.4$&$   -71.2$&$  -62.5$&$   -68.5$&$   -63.7$&$   -66.6$\\
$  2.8 $&$  -38.1( -31.6)\pm 8.1$&$   -61.3$&$  -51.0$&$   -55.9$&$   -52.4$&$   -54.3$\\
$  2.9 $&$  -34.5( -29.0)\pm 7.1$&$   -49.2$&$  -43.4$&$   -44.0$&$   -44.2$&$   -44.8$\\
$  3.0 $&$  -23.4( -19.3)\pm 6.0$&$   -39.4$&$  -37.4$&$   -37.0$&$   -34.9$&$   -37.2$\\
\noalign{\smallskip}
\hline
\end{tabular}
\end{center}
\small {\bf Note.} 
Genus values at a given threshold level of the SDSS galaxies in the BEST sample and 
five sets of the mock galaxies. The density fields were smoothed smoothing scale $R_G$ is $6.0\hmpc$.
The genus values of the SDSS galaxies are systematic bias-corrected and
the observed values are given in parentheses.
\end{table*}

{}


\end{document}